\input harvmac.tex
\input epsf


\def\ndt{\noindent}

\def\K3{{\bf K3}}
\def\journal#1&#2(#3){\unskip, \sl #1\ \bf #2 \rm(19#3) }
\def\andjournal#1&#2(#3){\sl #1~\bf #2 \rm (19#3) }

\def\bar{\overline}

\def\tilde{\widetilde}

\def\frac#1#2{{#1\over#2}}

\def\vev#1{\langle#1\rangle}
\def\d{\partial}

\def\inbar{\,\vrule height1.5ex width.4pt depth0pt}
\def\IC{\relax\hbox{$\inbar\kern-.3em{\rm C}$}}
\def\IR{\relax{\rm I\kern-.18em R}}
\def\IP{\relax{\rm I\kern-.18em P}}

%
%


%
\catcode`\@=11
\def\slash#1{\mathord{\mathpalette\c@ncel{#1}}}
\overfullrule=0pt
\def\AA{{\cal A}}

\def\GG{{\cal G}}

\def\MM{{\cal M}}
\def\NN{{\cal N}}

\def\VV{{\cal V}}

\def\ZZ{{\cal Z}}
\def\lam{\lambda}

\def\underrel#1\over#2{\mathrel{\mathop{\kern\z@#1}\limits_{#2}}}

\catcode`\@=12


%

\def\vev#1{\left\langle #1 \right\rangle}

\def\tr{{\rm tr}}

\def\exp{{\rm exp}}


\def \ov {\over}
\def \p {\partial}
\def \ha {{1 \ov 2}}
\def \al {\alpha}
\def \lam {\lambda}

\def \om {\omega}

\def \ep {\epsilon}

\def \apr {\alpha'}

\def\le{\left}
\def\ri{\right}

\def\IL{\relax{\rm I\kern-.18em L}}
\def\IH{\relax{\rm I\kern-.18em H}}
\def\IR{\relax{\rm I\kern-.18em R}}
\def\IC{\relax\hbox{$\inbar\kern-.3em{\rm C}$}}





\def\makeblankbox#1#2{\hbox{\lower\dp0\vbox{\hidehrule{#1}{#2}%
   \kern -#1
   \hbox to \wd0{\hidevrule{#1}{#2}%
      \raise\ht0\vbox to #1{}
      \lower\dp0\vtop to #1{}
      \hfil\hidevrule{#2}{#1}}%
   \kern-#1\hidehrule{#2}{#1}}}%
}%
\def\hidehrule#1#2{\kern-#1\hrule height#1 depth#2 \kern-#2}%
\def\hidevrule#1#2{\kern-#1{\dimen0=#1\advance\dimen0 by #2\vrule
    width\dimen0}\kern-#2}%
\def\openbox{\ht0=1.2mm \dp0=1.2mm \wd0=2.4mm  \raise 2.75pt
\makeblankbox {.25pt} {.25pt}  }

\def\bun#1/#2{\leavevmode
   \kern.1em \raise .5ex \hbox{\the\scriptfont0 #1}%
   \kern-.1em $/$%
   \kern-.15em \lower .25ex \hbox{\the\scriptfont0 #2}%
}

\def\opensquare{\ht0=3.4mm \dp0=3.4mm \wd0=6.8mm  \raise 2.7pt
\makeblankbox {.25pt} {.25pt}  }


\def\sector#1#2{\ {\scriptstyle #1}\hskip 1mm
\mathop{\opensquare}\limits_{\lower
1mm\hbox{$\scriptstyle#2$}}\hskip 1mm}

\def\tsector#1#2{\ {\scriptstyle #1}\hskip 1mm
\mathop{\opensquare}\limits_{\lower
1mm\hbox{$\scriptstyle#2$}}^\sim\hskip 1mm}



\lref\BasuMQ{
  P.~Basu, B.~Ezhuthachan and S.~R.~Wadia,
  arXiv:hep-th/0610257.
}

\lref\BasuPJ{
  P.~Basu and S.~R.~Wadia,
  Phys.\ Rev.\ D {\bf 73}, 045022 (2006)
  [arXiv:hep-th/0506203].
}

\lref\DeyDS{
  T.~K.~Dey, S.~Mukherji, S.~Mukhopadhyay and S.~Sarkar,
  arXiv:hep-th/0609038.
}

\lref\AlvarezGaumeJG{
  L.~Alvarez-Gaume, P.~Basu, M.~Marino and S.~R.~Wadia,
  arXiv:hep-th/0605041.
}

\lref\FuruuchiST{
  K.~Furuuchi,
  arXiv:hep-th/0608108.
}

\lref\SchnitzerXZ{
  H.~J.~Schnitzer,
  arXiv:hep-th/0612099.
}

\lref\HarmarkET{
  T.~Harmark, K.~R.~Kristjansson and M.~Orselli,
  arXiv:hep-th/0701088.
}

\lref\HarmarkTA{
  T.~Harmark and M.~Orselli,
  Phys.\ Rev.\ D {\bf 74}, 126009 (2006)
  [arXiv:hep-th/0608115].
}

\lref\HikidaQB{
  Y.~Hikida,
  arXiv:hep-th/0610119.
}

\lref\Hagedorn{
  R.~Hagedorn,
  Nuovo Cim.\ Suppl.\  {\bf 3}, 147 (1965).
}

\lref\Huang{
  K.~Huang and S.~Weinberg,
  Phys.\ Rev.\ Lett.\  {\bf 25}, 895 (1970).
}

\lref\sapa{
  B.~Sathiapalan,
  Phys.\ Rev.\ D {\bf 35}, 3277 (1987).
}

\lref\Fubini{
  S.~Fubini and G.~Veneziano,
  Nuovo Cim.\ A {\bf 64}, 811 (1969).
}

\lref\Tan{
  K.~H.~O'Brien and C.~I.~Tan,
  Phys.\ Rev.\ D {\bf 36}, 1184 (1987).
}

\lref\kogan{
  Y.~I.~Kogan,
  JETP Lett.\  {\bf 45}, 709 (1987)
  [Pisma Zh.\ Eksp.\ Teor.\ Fiz.\  {\bf 45}, 556 (1987)].
}

\lref\atic{
  J.~J.~Atick and E.~Witten,
  Nucl.\ Phys.\ B {\bf 310}, 291 (1988).
}

\lref\grossperryyaffe{
  D.~J.~Gross, M.~J.~Perry and L.~G.~Yaffe,
  Phys.\ Rev.\ D {\bf 25}, 330 (1982).
}

\lref\zwiebach{
  B.~Zwiebach,
  ``A first course in string theory,''
}

\lref\polchinski{
  J.~Polchinski,
  ``String theory. Vol. 1: An introduction to the bosonic string,''
}

\lref\polchik{
  J.~Polchinski,
  Commun.\ Math.\ Phys.\  {\bf 104}, 37 (1986).
}

\lref\kniznik{
  A.~A.~Belavin and V.~G.~Knizhnik,
  Sov.\ Phys.\ JETP {\bf 64}, 214 (1986)
  [Zh.\ Eksp.\ Teor.\ Fiz.\  {\bf 91}, 364 (1986)].
}

\lref\lawrence{
  M.~Kruczenski and A.~Lawrence,
  JHEP {\bf 0607}, 031 (2006)
  [arXiv:hep-th/0508148].
}

\lref\Hawkingpage{
  S.~W.~Hawking and D.~N.~Page,
  Commun.\ Math.\ Phys.\  {\bf 87}, 577 (1983).
}

\lref\dhoker{
  E.~D'Hoker and D.~H.~Phong,
  Rev.\ Mod.\ Phys.\  {\bf 60}, 917 (1988).
}

\lref\barbon{
  J.~L.~F.~Barbon and E.~Rabinovici,
  JHEP {\bf 0203}, 057 (2002)
  [arXiv:hep-th/0112173].
  J.~L.~F.~Barbon and E.~Rabinovici,
  arXiv:hep-th/0407236.
}

\lref\sundB{
  B.~Sundborg,
  Nucl.\ Phys.\ B {\bf 573}, 349 (2000)
  [arXiv:hep-th/9908001].
}

\lref\LiuVY{
  H.~Liu,
  ``Fine structure of Hagedorn transitions,''
  arXiv:hep-th/0408001.
}


\lref\MinW{
  O.~Aharony, J.~Marsano, S.~Minwalla, K.~Papadodimas and M.~Van Raamsdonk,
  Adv.\ Theor.\ Math.\ Phys.\  {\bf 8}, 603 (2004)
  [arXiv:hep-th/0310285].
}

\lref\raamsdonk{
  K.~Papadodimas, H.~H.~Shieh and M.~Van Raamsdonk,
  arXiv:hep-th/0612066.
}
\lref\AharonyBQ{
  O.~Aharony, J.~Marsano, S.~Minwalla, K.~Papadodimas and M.~Van Raamsdonk,
  Phys.\ Rev.\ D {\bf 71}, 125018 (2005)
  [arXiv:hep-th/0502149].
}
\lref\thooft{
  G.~'t Hooft,
  Nucl.\ Phys.\ B {\bf 72}, 461 (1974).
}

\lref\brigante{
  M.~Brigante, G.~Festuccia and H.~Liu,
  Phys.\ Lett.\ B {\bf 638}, 538 (2006)
  [arXiv:hep-th/0509117].
}

\lref\klebanov{
  D.~J.~Gross and I.~R.~Klebanov,
  Nucl.\ Phys.\ B {\bf 354}, 459 (1991);
\break
  I.~R.~Klebanov,
  ``String Theory In Two-Dimensions,''
  arXiv:hep-th/9108019.
}

\lref\LV{Landsman and Van Weert , Phys.\ Rept.\  }

\lref\sred{M.~Srednicki, Phys.\ Rev.\ E {\bf 50} 888 (1994).}

\lref\ccgi{G.~Casati1, B.~V.~Chirikov1,2, I.~Guarneri1 and
F.~M.~Izrailev, cond-mat/9607081.}

\lref\fcic{Y.~V.~Fyodorov, O.~A.~Chubykalo, F~.M~Izrailev and
G.~Casati, Phys. \ Rev. \ Lett. \ {\bf 76} 1603 (1996).}

\lref\ConstableRC{
  N.~R.~Constable and F.~Larsen,
  ``The rolling tachyon as a matrix model,''
  JHEP {\bf 0306}, 017 (2003)
  [arXiv:hep-th/0305177].
}

\lref\dineetal{
  M.~Dine, E.~Gorbatov, I.~R.~Klebanov and M.~Krasnitz,
  JHEP {\bf 0407}, 034 (2004)
  [arXiv:hep-th/0303076].
}

\lref\McClainID{
  B.~McClain and B.~D.~B.~Roth,
  ``MODULAR INVARIANCE FOR INTERACTING BOSONIC STRINGS AT FINITE TEMPERATURE,''
  Commun.\ Math.\ Phys.\  {\bf 111}, 539 (1987).
 }

\lref\AlvarezGaumeFV{
  L.~Alvarez-Gaume, C.~Gomez, H.~Liu and S.~Wadia,
  Phys.\ Rev.\ D {\bf 71}, 124023 (2005)
  [arXiv:hep-th/0502227].
}


\lref\adscft{
  J.~M.~Maldacena,
  ``The large N limit of superconformal field theories and supergravity,''
  Adv.\ Theor.\ Math.\ Phys.\  {\bf 2}, 231 (1998)
  [Int.\ J.\ Theor.\ Phys.\  {\bf 38}, 1113 (1999)]
  [arXiv:hep-th/9711200].
\break
  S.~S.~Gubser, I.~R.~Klebanov and A.~M.~Polyakov,
  ``Gauge theory correlators from non-critical string theory,''
  Phys.\ Lett.\ B {\bf 428}, 105 (1998)
  [arXiv:hep-th/9802109].

\break

  E.~Witten,
  ``Anti-de Sitter space and holography,''
  Adv.\ Theor.\ Math.\ Phys.\  {\bf 2}, 253 (1998)[arXiv:hep-th/9802150].

}

\lref\GomezReinoBQ{
  M.~Gomez-Reino, S.~G.~Naculich and H.~J.~Schnitzer,
  JHEP {\bf 0507}, 055 (2005)
  [arXiv:hep-th/0504222].
}

\lref\SchnitzerQT{
  H.~J.~Schnitzer,
  Nucl.\ Phys.\ B {\bf 695}, 267 (2004)
  [arXiv:hep-th/0402219].
}

\lref\SpradlinPP{
  M.~Spradlin and A.~Volovich,
  Nucl.\ Phys.\ B {\bf 711}, 199 (2005)
  [arXiv:hep-th/0408178].
}




\Title{\vbox{\baselineskip12pt \hbox{hep-th/0701205}
\hbox{MIT-CTP-3803} \hbox{NSF-KITP-07-06}
}}%
 {\vbox{\centerline{Hagedorn divergences and tachyon potential}
 } }

\smallskip
\centerline{Mauro Brigante$^1$, Guido Festuccia$^{1}$ and Hong
Liu$^{1,2}$ }

\medskip

\centerline{$^1$ {\it  Center for Theoretical Physics}}
\centerline{\it Massachusetts Institute of Technology}
\centerline{\it Cambridge, Massachusetts, 02139}

\medskip

\centerline{$^2$ {\it Kavli Institute for Theoretical Physics}}
\centerline{\it University of California} \centerline{\it Santa
Barbara, CA 93106-4030}

\smallskip

\smallskip

\smallskip

\vglue .3cm

\bigskip
\noindent

We consider the critical behavior for a string theory near the
Hagedorn temperature. We use the factorization of the worldsheet
to isolate the Hagedorn divergences at all genera. We show that
the Hagedorn divergences can be resummed by introducing double
scaling limits, which smooth the divergences. The double scaling
limits also allow one to extract the effective potential for the
thermal scalar. For a string theory in an asymptotic anti-de
Sitter (AdS) spacetime, the AdS/CFT correspondence implies that
the critical Hagedorn behavior and the relation with the effective
potential should also arise from the boundary Yang-Mills theory.
We show that this is indeed the case. In particular we find that
the free energy of a Yang-Mills theory contains ``vortex''
contributions at finite temperature. Yang-Mills Feynman diagrams
with vortices can be identified with contributions from boundaries
of moduli space on the string theory side.

 \Date{January 2007}


\bigskip



\newsec{Introduction}

Since the early days of string theory, it was observed that the free
string spectrum has a density of states which grows exponentially
with energy, and that the partition function $Z= e^{-\beta H}$ of a
free string gas at a temperature $T={1 \ov \beta}$  would diverge
when $T$ is greater than some critical value
$T_H$~\refs{\Hagedorn,\Huang,\Fubini}. The Hagedorn divergence
occurs for all known (super)string theories with spacetime
dimensions greater than two. The physical meaning of the critical
temperature $T_H$ and of the divergence has been a source of much
discussion since then.

In the late eighties, a few important observations were made which
suggested that the Hagedorn divergence signals a phase transition,
analogous to the deconfinement transition in
QCD~\refs{\sapa,\kogan,\Tan,\atic}. At the Hagedorn temperature
$T_H$ the lowest winding modes (with winding $\pm 1$) around the
periodic Euclidean time direction become marginal operators in the
worldsheet conformal field theory~\refs{\sapa,\kogan,\Tan}.
Sathiapalan and Kogan~\refs{\sapa,\kogan} argued that above the
Hagedorn temperature, the winding modes would condense in a
fashion similar to the Kosterlitz-Thouless transition in the X-Y
model and the worldsheet theory will flow to a new infrared fixed
point. From the spacetime point of view, these winding modes (with
winding $\pm 1$) correspond to a complex scalar field $\phi$
living in one fewer spacetime dimension (i.e. not including
Euclidean time). Near the Hagedorn temperature, the spacetime
effective potential for $\phi$ can be written in a form
 \eqn\ejro{
 V = m^2_\phi (\beta) \phi^* \phi + \lam_4 g_s^2 (\phi^* \phi)^2 +
 \lam_6 g_s^4 (\phi^* \phi)^3 +
 \cdots, \qquad m^2_\phi (\beta) \propto T_H - T \ .
 }
If $\lam_4$ is positive (negative), the phase transition would be
second order (first order). In~\refs{\atic} Atick and Witten
argued that for a string theory in asymptotic flat spacetime the
transition should be first order\foot{That the transition is of
first order can also be argued from the non-perturbative
instability of the thermal flat spacetime discovered
in~\refs{\grossperryyaffe}.} (i.e. $\lam_4 < 0$) due to the
coupling of the thermal scalar to the dilaton.

While the one-loop Hagedorn divergence has been extensively
discussed in the past~(see e.g.~\refs{\zwiebach,\polchinski} for
reviews), Hagedorn divergences from higher genus amplitudes have
been investigated rather little. In this paper we use a
factorization argument to extract Hagedorn divergences for higher
genus amplitudes. We show that they can be resumed by introducing
various double scaling limits, which smooth the divergences. The
double scaling limits also allow one to extract the effective
potential \ejro\ to arbitrary high orders. That a double scaling
limits might exist for higher genus Hagedorn divergences was
speculated earlier in~\refs{\LiuVY} and further discussed
in~\refs{\AlvarezGaumeFV} in a toy model motivated from AdS/CFT.

Our discussion further highlights that Hagedorn divergences signal
a breakdown of string perturbation theory due to appearance of
massless modes and do not imply a limiting temperature for string
theory~\refs{\sapa,\kogan,\atic}.

The discussion of this paper will be rather general, e.g.
applicable to string theories in asymptotic anti-de Sitter (AdS)
spacetime. The AdS/CFT correspondence then implies that the
critical Hagedorn behavior from high genera and the relation with
the effective potential should also arise from Yang-Mills
theories. We show that this is indeed the case. In particular we
find that the free energy of Yang-Mills theory contains ``vortex''
contributions at finite temperature. Yang-Mills Feynman diagrams
with vortices can be identified with contributions from the
boundary of the moduli space on the string theory side.

The plan of the paper is as follows. In section 2 we first review
the one-loop result and discuss the physical set-up of our
calculation. We then extract the critical Hagedorn behavior from
higher genus amplitudes and show that one can find terms in \ejro\
by defining suitable double scaling limits. In section 3 we turn to
Yang-Mills theory. We discuss the structure of the large $N$
expansion for the partition function of a Yang-Mills theory at
finite temperature and isolate the critical Hagedorn behavior. We
conclude in section 4 with a discussion of some physical
implications.

\newsec{High-loop Hagedorn divergences in perturbative string theory}

\subsec{Review of one-loop divergence and set-up}

Consider a string theory consisting of a compact CFT times
$\IR^{1,d}$. The one-loop free energy of the system at a finite
temperature can be computed by the torus path integral with a
target space in which the Euclidean time direction is compactified
with period $\beta = {1 \ov T}$ and with anti-periodic boundary
condition for spacetime fermions~\refs{\polchik}. The Hagedorn
singularity appears when the lowest modes with winding $\pm 1$
around the compactified time direction become
massless~\refs{\Tan,\sapa,\kogan}. More explicitly, the mass
square can be written as
 \eqn\ejsp{
 m^2_\phi (\beta) =  \le({\beta \ov 2 \pi \apr}\ri)^2  - c_0 \equiv
   \le({\beta \ov 2 \pi \apr}\ri)^2 - \le({\beta_H \ov 2 \pi \apr}\ri)^2
 }
where the first term is the winding contribution and $c_0$ is the
zero point energy of the string~(in the winding sector). The
second equality of \ejsp\ should be considered as a definition of
the Hagedorn temperature. From \ejsp, $m^2_\phi (\beta) \to 0$ as
$\beta \to \beta_H$ and becomes tachyonic when $\beta < \beta_H$.
In spacetime, the  winding $\pm 1$ modes correspond to a complex
scalar field $\phi$ living in one fewer spacetime dimension (i.e.
spatial part of the spacetime), which is often called the thermal
scalar in the literature. We will follow this terminology below.
We will also collectively call modes with general winding numbers
(and no internal excitations) winding tachyons. Equation \ejsp\
applies to both bosonic and superstring theories with possibly
different $c_0$ for different theories.

 The critical behavior of the one-loop free energy $F_1$ as $\beta \to \beta_H$
is controlled by that of the thermal scalar
 \eqn\rus{
 F_1 = -2 \times \ha \log (-\nabla^2 + m_\phi^2 (\beta)) + F_{finite},
 \qquad \beta \to \beta_H
 }
where $\nabla^2$ is the Laplacian on the {\it spatial} manifold.
If the gap of $\nabla^2$ along the compact CFT directions is
bigger than $m_\phi^2 (\beta)$, the singular part of \rus\ can be
further written as
 \eqn\fhds{\eqalign{
 F_{1} & \propto  -  \int {d^d k \ov (2 \pi)^d} \log (k^2 + m_\phi^2
 (\beta))+ \cdots
 \cr
 & \propto \cases{(m^2_\phi(\beta))^{d \ov 2} & $d \;\; {\rm odd}$ \cr
             (m^2_\phi (\beta))^{d \ov 2} \log m^2_\phi (\beta) & $d \;\; {\rm even}$ \cr
             } \
 }}
$F_{1}$ has a branch point singularity at $m^2_\phi (\beta) =0$
for all $d$.  In particular, for $d=0$ it is logarithmically {\it
divergent} as $\beta \to \beta_H$
 \eqn\fkd{
 F_1 = - \log (\beta - \beta_H) + {\rm finite} \ .
 }

The above discussion should also apply to a static curved
spacetime, for example, an AdS spacetime, even though an explicit
computation of
the one-loop free energy is often not possible. 
For an AdS spacetime, since the Laplacian has a mass gap, we
expect the free energy for a thermal gas of AdS strings should
behave as \fkd\ when the Hagedorn temperature is approached~(see
e.g.~\refs{\lawrence} for further discussion).

In this paper we will focus our discussion on $d=0$ or more
generally those spacetimes (including AdS) in which \fkd\ is
satisfied, for the following reasons:

\item {1.}  The thermal ensemble cannot be defined in an uncompact
asymptotically flat spacetime due to Jeans instability. To make
the canonical ensemble well defined, an Infrared (IR) cutoff is
needed. One particularly convenient (and well-defined) IR
regulator is to introduce a small negative cosmological
constant\foot{In an asymptotic AdS spacetime it is possible to
define a canonical ensemble in the presence of gravity, as
discovered by Hawking and Page~\refs{\Hawkingpage}. Hawking and
Page also found that the system undergoes a first order phase
transition at a temperature $T_{HP}$ from a thermal gas in AdS to
a stable black hole. Treating an AdS spacetime with a small
cosmological constant as an IR regularization of the flat
spacetime, it is natural to identify the first order phase
transition argued by~\refs{\atic} with the Hawking-Page
transition. Note that the flat space limit, which corresponds to
keeping $g_s$ small, but fixed and taking the curvature radius of
AdS to infinity, is rather subtle. In this limit the stable black
hole phase in AdS disappears and the Jeans instability should
develop at a certain point. Also note that in the flat space
limit, the Hawking-Page temperature goes to zero, which is
consistent with the observation that a hot flat spacetime is
non-perturbatively unstable at any nonzero
temperature~\refs{\grossperryyaffe}.}. For our discussion below
the precise nature of such a regulator will not be important as
far as it makes the thermal ensemble well defined. Such IR
regulators introduce a gap in the Laplacian $\nabla^2$, which will
be kept fixed in the limit $T \to T_H$ and thus will be greater
than $m^2_{\phi}$ when the temperature is sufficiently close to
$T_H$.

\item{2.} The Hagedorn singularity is sharpest at $d=0$. While the
free energy is singular at $\beta = \beta_H$ for all dimensions,
it is divergent only for $d=0$.

The logarithmic divergence of \fkd\ at $\beta \to \beta_H$ implies
that the string perturbation theory breaks down {\it before}
$\beta = \beta_H$ is reached. Thus it is not sufficient to
consider only the one-loop contribution to the free energy and
higher genus contributions could be important. Below we will show
that as $\beta \to \beta_H$, it is necessary to resum the string
perturbation theory to all orders. We will then show that one can
extract the spacetime effective action for the thermal scalar from
the resumed series and that the divergences are smoothed out.

When $\lam_4$ in \ejro\ is negative, i.e. when the transition is
first order, there exists a lower temperature $T_c < T_H$, at
which the thermal gas of strings becomes metastable. At a
temperature $T_c < T < T_H$, the thermal gas is still
perturbatively stable. Here we are interested in probing the
critical behavior in the metastable phase (or superheated phase)
as $T \to T_H$ from below.

\subsec{Higher loop divergences}

We now examine higher loop divergences as $\beta \to \beta_H$. For
simplicity we will restrict our discussion to bosonic strings. We
expect the conclusion to hold for superstring theories as well.

 The
genus-$g$ contribution $F_g$ to the free energy is obtained by
integrating the single string partition function on a genus-$g$
surface over the moduli space $\MM_g$ of such surfaces. The
potentially divergent contributions to $F_g$ arise from the
integration near the boundary of the moduli space.

\ifig\Degen{An example of a degenerate genus-$6$ Riemann surface.
 Each blob represents a surface
of certain genus and thin lines connecting blobs
 represent pinched cycles.} {\epsfxsize=8cm \epsfbox{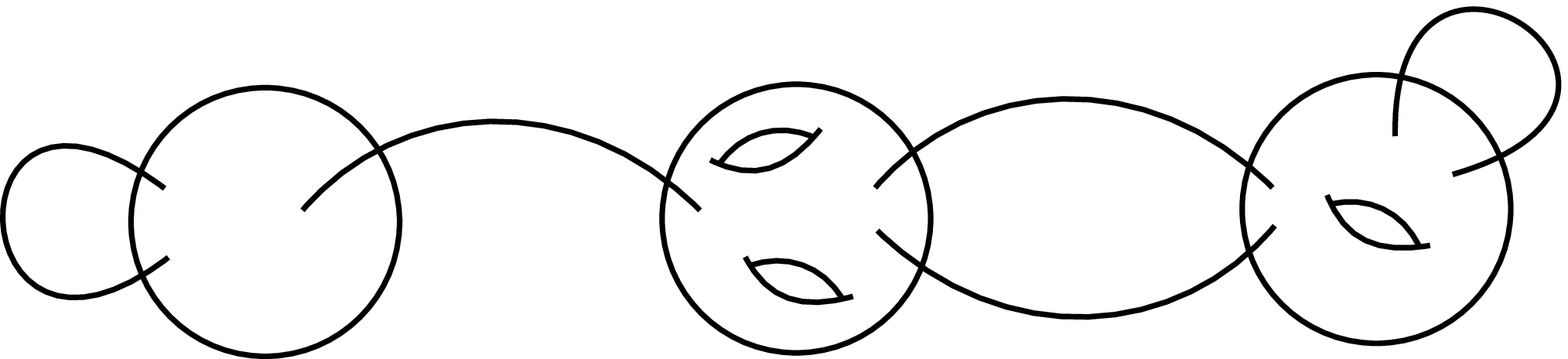}}

\ifig\genTwo{Degenerate limits of a genus-$2$ Riemann surface.
 } {\epsfxsize=10cm \epsfbox{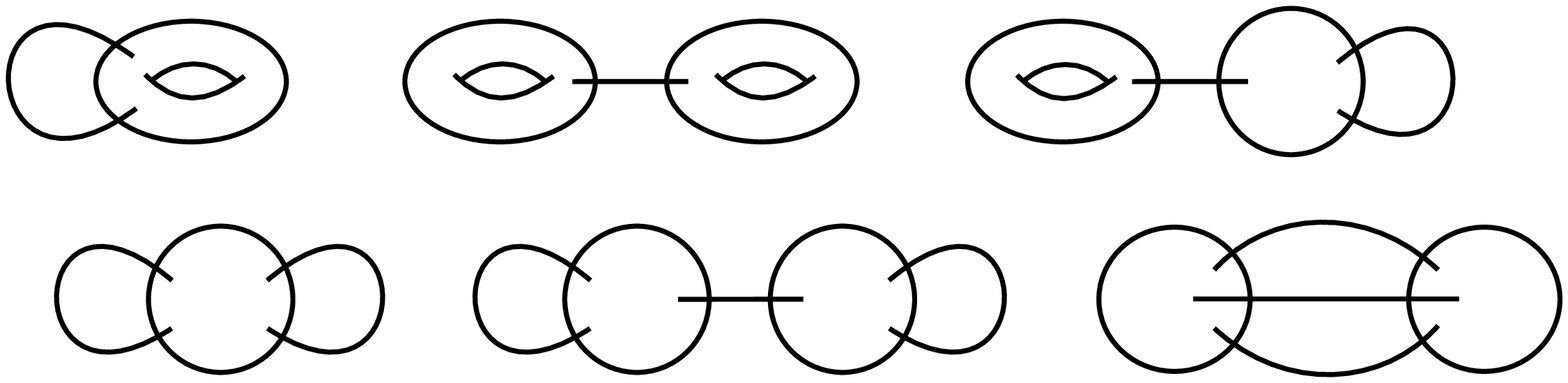}}

The boundary $\Delta_g$ of $\MM_g$ is where a Riemann surface
degenerates, which can be described by pinching cycles on the
surface (for reviews see
e.g.~\refs{\dhoker,\kniznik,\polchinski}).
 There are two types of basic degenerations depending on whether the pinched
 cycle is homologous to zero or not. If the pinched cycle is homologous to
 zero, a surface of genus $g$
degenerates into two surfaces of genus $g_1$ and $g_2$ ($g=g_1 +
g_2$) which are joined together at a point. If the pinched cycle is
not homologous to zero, a genus $g$ surface degenerates into a
surface of genus $g-1$ with two points glued together.  One can
pinch more than one cycle at the same time as far as they do not
intersect with each other. On a genus $g$ surface, the maximal
number of nonintersecting closed geodesics is $3g-3$, so one can
pinch at most $3g-3$ cycles at the same time. See \Degen\ and
\genTwo\ for examples of degenerate limits.

Let us now examine the contribution to $F_g$ from boundaries of
moduli space. The pinching of a Riemann surface can be described
in terms of cutting open the path integral on the surface. The
pinching is a local operation and so is cutting the path integral
(other than possible constraints from the zero mode integration).
We follow the standard procedure as described
in~\refs{\polchinski}. One has
 \eqn\fhs{
 \vev{1}_{g} = \sum_{i} q^{h_i} \bar
 q^{\tilde h_i} \vev{ \AA_i (z_1)}_{{g_1}}
  \vev{\AA_i (z_2)  }_{{g_2}}
 }
and
  \eqn\rjs{
 \vev{1}_{g} = \sum_{i} q^{h_i} \bar
 q^{\tilde h_i} \, \vev{\AA_i (z_1) \AA_i (z_2)  }_{{g-1}}
 }
for the two types of basic degenerations, where $\vev{\cdots}_g$
denote worldsheet correlation functions on a genus $g$ surface and
$i$ sums over a complete set of intermediate states. $q$ can be
considered as the complex coordinate transverse to the boundary with
$q \to 0$ corresponding to the degeneration limit. Integration of
\fhs\ and \rjs\ near $q \to 0$ yields the propagator
  \eqn\rkna{
 G  
 = \sum_i { 8 \pi
  \ov \apr (-\nabla^2 + m_i^2)} \ .
 }

The contribution to the free energy from boundaries of moduli
space can be extracted from diagrams like the ones in \Degen\ and
 \genTwo.
 One can treat blobs (representing surfaces of certain
 genus with some insertions) as effective vertices
 and thin lines (pinched cycles) as  propagators.
For $\beta \to \beta_H$ and assuming that the spatial Laplacian
operator $-\nabla^2$ has a gap, then the propagator \rkna\ for a
pinched cycle is potentially dominated by that of the thermal
scalar\foot{Note that it is not immediately obvious that the
thermal scalar (or other winding modes along the Euclidean time
direction) appears in the intermediate states from the point of
view of calculating the free energy of a finite temperature string
gas, since they do not correspond to spacetime physical states.
Indeed in the one-loop calculation, they appear only after a
modular transformation. However, it is clear that they should
appear in the intermediate states from the point of view that we
are working with a string theory compactified on a circle with
anti-periodic boundary condition for fermions.},
  \eqn\nsa{
 G \approx {8 \pi  \ov \apr m_\phi^2 (\beta)} + {\rm finite}
  \propto {1 \ov \beta - \beta_H} + \cdots , \quad \beta \to
  \beta_H
  }
Since one can pinch at most $3g-3$ cycles at the same time,
naively we may conclude from \nsa\ that $F_g$ diverges as
 $
  {1 \ov (\beta - \beta_H)^{3g-3}}
  $
 for $g \geq 2$ as $\beta \to \beta_H$.
However, there are global constraints due to winding number
conservation at each blob of \Degen\ and \genTwo. As a result, not
all propagators can have the nearly-massless thermal scalar
propagating through them. We will now show that the most divergent
terms at genus $g$ are proportional to
  \eqn\rjj{
  {1 \ov (\beta - \beta_H)^{2g-2}} \qquad g \geq 2 \ .
  }

Let us consider a generic degenerate limit of a genus $g$ surface
as shown for example in \Degen. Denote $V^{(n,m)}$ the number of
vertices with genus $n$ and $m$ insertions. Then the total number
$L$ of pinched cycles (propagators) and the genus $g$ of the whole
surface can be written as
 \eqn\ris{
 2L = \sum_{n,m} m V^{(n,m)}, \qquad g = 1 + \sum_{n,m} \le({m \ov 2} +n -1\ri) V^{(n,m)}
 \ .
 }
The second equation of \ris\ can be obtained from the degenerate
rules stated earlier. Alternatively, one can associate each
insertion with a factor of $g_s$ and the total power of $g_s$
should be $2(g-1)$. It is also convenient to introduce
 \eqn\ruur{
V = \sum_{n,m} V^{(n,m)}, \qquad g_a = \sum_{n,m} n V^{(n,m)},
 }
where $V$ is the total number of vertices, $g_a$ is the apparent
genus of the diagram (i.e. the sum of the genus of each vertex).
Equations \ris\ and \ruur\ lead to
 \eqn\fjkr{
 L - ( V-1) = g - g_a  \ .
 }
Since winding numbers carried by propagators have to be conserved
at each vertex, equation \fjkr\ implies that the total number of
independent windings in a diagram is $g - g_a$. The maximal number
of independent windings among different degenerate limits is then
$g$, in which cases each vertex has the topology of a sphere.

From \ris\ and \ruur, we also have
 \eqn\orjw{
 V = \ha L - {1 \ov 4} \sum_{n,m} (m-4) V^{(n,m)}
 }
and  \orjw\ and \fjkr\ lead to
 \eqn\runs{
 L = 2 (g-1) - \sum_{n,m} (2n + {m \ov 2} -2) V^{(n,m)} \ .
 }
Equation \runs\ implies that the maximal number of propagators
(pinched cycles) in a degenerate limit is indeed $3g-3$, obtained
when only $V^{(0,3)}$ is nonzero. However, it is impossible to have
all $3g-3$ propagators to be divergent at the same time, i.e. to
have all windings to be $\pm 1$, since by winding number
conservation if the windings of two of the propagators coming out of
a 3-point vertex are $\pm 1$, then the third one can only be $0, \pm
2$.

\ifig\genusT{Two possible degenerate limits of a genus-$3$ Riemann
surface which give rise to most divergent contributions. Each
propagator has the thermal scalar running through it.}
{\epsfxsize=10cm \epsfbox{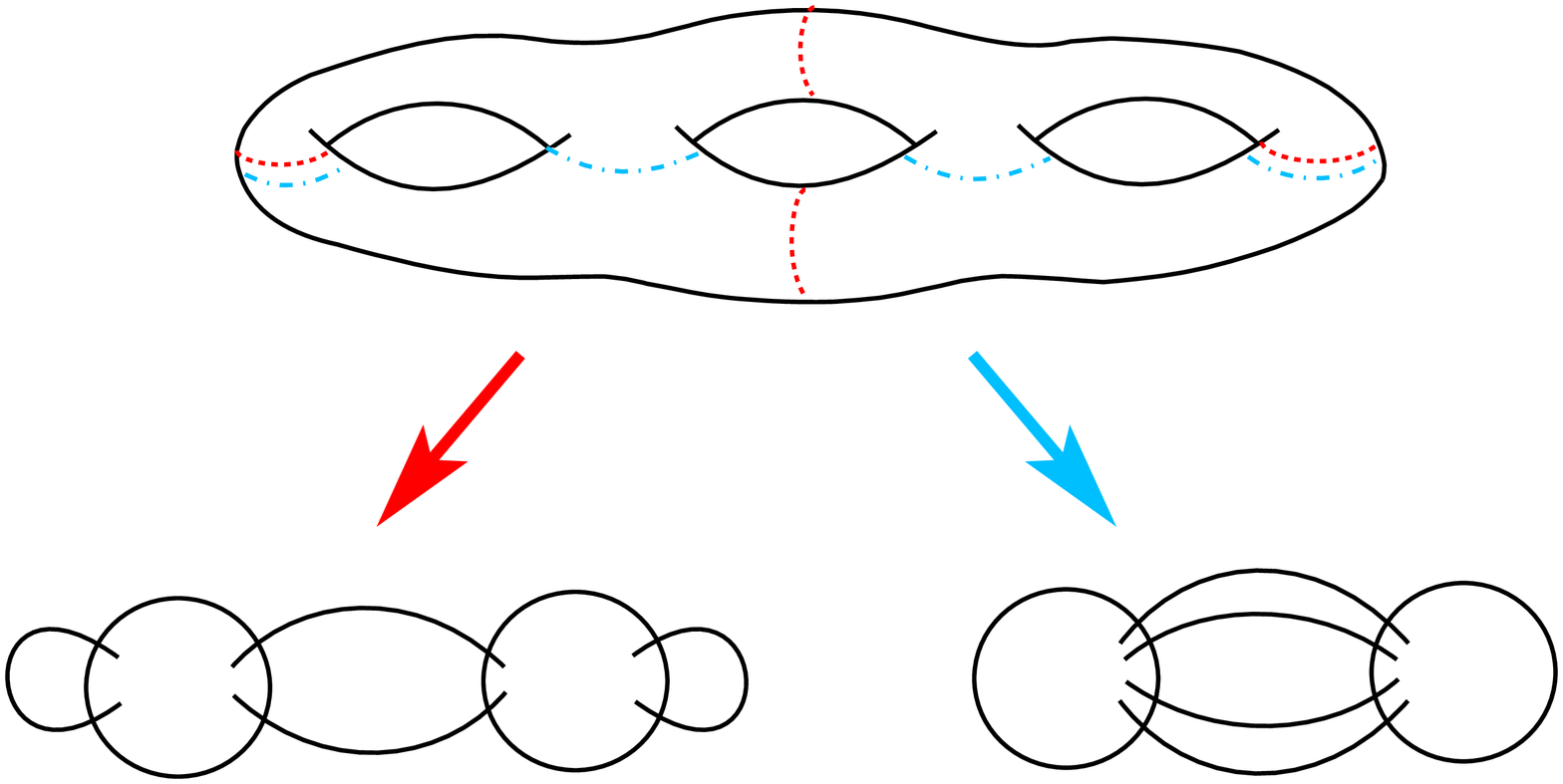}}

Since at least one of the propagators going out of a 3-point vertex
must have winding $|w|\neq 1$, if our purpose is to find the {\it
maximum} number of propagators that can have $w= \pm 1$, one can
ignore such a propagator. This implies we  only need to consider
those degenerate limits in which effective vertices have at least
four insertions, i.e. $m \geq 4$. In the absence of $V^{(0,3)}$,
equation \runs\ implies that
 \eqn\rjdj{
 L \leq 2 (g-1)
 }
 where the equality holds when
 \eqn\shal{
 V^{(0,4)} \neq 0, \qquad {\rm otherwise} \;\; V^{(n,m)} = 0 \ .
 }
Thus we have proven that  the most divergent term is of the form
\rjj. See \genusT\ for degenerations which give rise to the most
divergent contributions at genus $3$.

To summarize, the most divergent contributions at each genus have
the following diagrammatic structure:

\item{1.} Each vertex has the topology of a sphere and has four
winding tachyon operator insertions with winding numbers
$1,1,-1,-1$ respectively. The total number of vertices in a genus
$g$ diagram is $g-1$. The path integral over each vertex gives
rise to an effective coupling
 \eqn\lamST{
 g_s^2 \tilde \lambda_4 = \vev{\VV_{+1} (0) \VV_{+1} (1) \VV_{-1} (\infty) \int d^2 z \, \VV_{-1} (z) }_{S^2}
 }
Note that at $\beta = \beta_H$, the vertex operators $\VV_{\pm 1}$
for the thermal scalar are marginal and \lamST\ is well defined.
Also $\tilde \lam_4$ is $g_s$-independent.

\item{2.} The propagators are given by that of the winding tachyon
\nsa. The total number of propagators is $2 (g-1)$.

\ndt Thus the most divergent contribution to the free energy at
genus-$(n+1)$ has the form
 \eqn\fjrl{
 a_n g_s^{2n} \tilde \lam^n_4
 \le({8 \pi  \ov \apr m_\phi^2 (\beta)}\ri)^{2n} \propto {g_s^{2n}
 \ov (\beta - \beta_H)^{2n}}
 }
where  $a_n$ is a combinatoric numerical factor depending on the
specific geometric structure of boundaries of moduli space.
Determining these numerical factors from direct worldsheet
computation is a rather complicated mathematical question, which
goes beyond the scope of this paper. In the next subsection we
will determine them using an indirect argument.

 \subsec{Double scaling limits and the effective thermal scalar
action}

In the last subsection we showed that the leading order Hagedorn
divergences at all loop orders can be written as
 \eqn\rjsl{\eqalign{
 F_{sing} & =- \log (\beta-\beta_H) + a_1  {\lam_4  g_s^2 \ov
  m_\phi^4}
  + \cdots +  a_n \le({g_s^{2} \lam_4 \ov  m_\phi^4} \ri)^n
 + \cdots \
  }}
with
 \eqn\jrjp{
 \lam_4 = \tilde \lam_4 \le({8 \pi \ov \apr} \ri)^2 , \qquad m_{\phi}^2
 \approx {\beta_H \ov 2 \pi^2 \apr} (\beta - \beta_H)  \ .
 }
Equation \rjsl\ suggests a double scaling limit
 \eqn\rkjk{
 \beta - \beta_H \to 0 , \qquad g_s \to 0, \qquad { \beta -
 \beta_H \ov g_s} = {\rm finite}
 }
in which case all higher order terms in the series become equally
important and we need to be resumed.

How do we interpret the free energy $F$ obtained by resuming the
series? A clue comes from the structure of the degenerate diagrams
summarized at the end of the last subsection, which resemble the
Feynman diagrams of a $ |\phi|^4$ theory (see e.g. \genusT).
Indeed the free energy of a $ |\phi|^4$ theory gives an asymptotic
expansion which is precisely of the form \rjsl\ with specific
values for the numerical coefficients $a_n$. Given that string
theory should reduce to a field theory in the low energy limit,
and that here we are essentially isolating an effective theory for
the nearly-massless thermal scalar, it is natural to conjecture
that \rjsl\ can be written as
 \eqn\djsa{\eqalign{
 F_{sing}  & = \log \int d \phi d \phi^* \, e^{-  m_\phi^2 \phi \phi^* - g_s^2 \lam_4 (\phi
\phi^*)^2} \cr
 & = - \log (\beta - \beta_H) - 2 {g_s^2 \lam_4 \ov  m_\phi^4} + 10
 {g_s^4 \lam^2_4 \ov  m_\phi^8} + \cdots
 }}
with $\phi$ is a c-number. Equation \djsa\ determines $a_n$ to all
orders uniquely and implies the following effective potential for
the thermal scalar
 \eqn\effad{
  V =   m^2_\phi \phi \phi^* + \lam_4 g_s^2 (\phi \phi^*)^2 +
  \cdots  \ .
 }
In the next section we will show that the effective action \effad\
and \djsa\ arises from the critical behavior of Yang-Mills theories
near the Hagedorn temperature. Using AdS/CFT this would serve as a
proof of \djsa\ for string theories in an asymptotic AdS spacetime.
Furthermore, since the factors $a_n$ in \fjrl\ and \rjsl\ depend
only on the mathematical structure of the  moduli space of Riemann
surfaces and not on the specific string theory, the Yang-Mills
theory results serve as an indirect proof of \djsa.

It is clear from equation~\djsa\ that Hagedorn divergences at each
genus order in \rjsl\ simply signal breakdown of the asymptotic
expansion in $g_s$ due to that $\phi$ becomes massless. The
$m_\phi \to 0$ limit is apparently smooth in the resumed integral
expression \djsa. When $\lam_4$ is positive, i.e. when the
transition is second order, the integral \djsa\ is finite and
non-perturbatively defined. For negative (or zero) $\lam_4$, i.e.
when the transition is first order, the integral \djsa\ is not
defined non-perturbatively and higher order terms in the effective
potential are needed. In either cases the $m_\phi \to 0$ limit is
well-defined.

Equation \djsa\ implies that $a_n \sim n!$ for $n$ large. This is
in contrast with the $(2n)!$ growth of the asymptotic behavior for
the full free energy. Here we are only looking at contributions
from boundaries of moduli space, which accounts for the slower
growth. When $\lam_4 < 0$, one can formally resum the series
\rjsl\ or the second line of \djsa\ using Borel resumation and one
finds that the free energy contains an imaginary part of the form
$e^{-{1 \ov g_s^2}}$ due to the $n!$ growth of $a_n$. Such an
imaginary part can be interpreted as the tunnelling rate from the
metastable thermal string gas to the true non-perturbative minimum
(see also discussion in~\refs{\AlvarezGaumeFV}).

Here we have been focusing  on the lowest spacetime
mode\foot{Recall that we assume that the Laplacian of the
spacetime manifold has a mass gap.} of the thermal scalar, which
gives the most divergent contribution to the free energy. This
explains the finite-dimensional integral in~\djsa. From general
covariance it seems natural to generalize \effad\ to include
derivatives
 \eqn\rhjH{
 S  = \int d^d x \sqrt{g} \, \le(  |\p \phi|^2 + m_\phi^2 \phi \phi^*
 + \lam_4 g_s^2 (\phi \phi^*)^2  + \cdots \ri) \ .
 }
where $d^d x$ integrates over the spatial directions\foot{Note
that for an AdS with a small negative cosmological constant,
\rhjH\ applies to regions in the interior of the spacetime, since
in AdS $g_{tt}$ component of the metric is nontrivial and the
thermal scalar always has a large mass near the boundary.}.

Let us now consider the generalization of the above double scaling
argument to extract higher orders terms in \effad. From
equation~\ris\ the leading contribution of a generic degenerate
surface to the free energy can be written in the form
  \eqn\contrib{
 {g_s^{2g-2} \ov (\beta-\beta_H)^{L}} ={g_s^{\sum_{n,k} V^{(n,2k)}
(2n+2k-2)} \ov (\beta-\beta_H)^{ \sum_{n,k} k V^{(n,2k)}}}
 }
where in writing down \contrib\ we have assumed that
 all propagators in a degenerate diagram carry winding numbers\foot{If there is
 a propagator carrying a
winding number other than $\pm 1$, we can treat the two vertices
connected by this propagator as a single effective vertex. Keeping
doing this we obtain a degenerate diagram whose propagators only
carry winding numbers $\pm 1$.} $\pm 1$ and that each vertex
contains an even number of insertions $m=2k, k=2,3, \cdots$, due to
winding number conservation. Now consider the double scaling limit
 \eqn\fjsN{
 {(\beta-\beta_H) \ov g_s^a} = {\rm finite}, \qquad g_s \to 0
 }
under which \contrib\ is proportional to $g_s^K$ with $K$ given by
 \eqn\contribB{
 K  = \sum_{n=0}^\infty \sum_{k=2}^\infty V^{(n,2k)} (2n+2k-2 - k a ) \ .
 }
For $a < 1$, we always have $K > 0$ for any choice of $V^{(n,2k)}$.
At $a=1$, we get $K=0$ for diagrams with $V^{(0,4)} \neq 0$ only
while $K
>0$ for all other diagrams. In the double scaling limit \fjsN\
only the contributions of diagrams with $K=0$ survive. These are the
most divergent contributions we isolated in \rjsl\ and lead to the
effective action \effad. Now let us set by hand $\lam_4 =0$, then in
\contribB, $V^{(0,4)}=0$. The most divergent contributions in the
remaining diagrams are isolated by taking $a = {4 \ov 3}$, at which
$K=0$ for diagrams with $V^{0,6} \neq 0$ only and $K >0$ for all the
rest. In other words now the most divergent contributions to the
free energy can be written as
  \eqn\uene{
 F  = - \log (\beta-\beta_H) + {c_1 g_s^2 \ov (\beta-\beta_H)^{3 \ov 2}}
  + \cdots +  {c_n g_s^{2n} \ov (\beta - \beta_H)^{3n \ov 2}} + \cdots
 }
 which implies the effective
 potential
  \eqn\rmbs{
   V =   m_\phi^2 \phi \phi^* + \lam_6(\phi \phi^*)^3 +
  \cdots
  }
where $\lam_6$ is related to the genus-0 six-point function of the
vertex operators for the thermal scalar on the worldsheet. Now
restoring $\lam_4$ and combining \djsa\ and \rmbs\ we would conclude
that the effective potential can be written as
 \eqn\Mmbs{
   V =   m_\phi^2 \phi \phi^* + \lam_4 (\phi \phi^*)^2 + \lam_6 (\phi \phi^*)^3 +
  \cdots
  }
 The same procedure can then be repeated to the next order by first setting $\lam_4$
and $\lam_6$ to zero and then extracting the most divergent terms
in the remaining diagrams. One can continue this to arbitrary
orders in $(\phi \phi^*)^n$ and we find the effective
potential\foot{Note that the procedure is not well adapted to
resum divergences due to vertices with genus $n \geq 1$. From
\contribB, to have $K=0$ for $n =1$, we need $a=2$, in which case
all genus $1$ vertices with arbitrary number of insertions
contribute equally. To have $K=0$ for $n >1$, we need $a>2$, then
from \contribB, diagrams with large $k$ become more dominant
regardless of the value of $n$.}
 \eqn\Mmbsdg{
  V =  m^2_\phi \phi \phi^* + \sum_{k=2}^{\infty} \lam_{2k} g_s^{2k-2}
  (\phi \phi^*)^k +
  \cdots
  }
The $\lambda_{2k}$ term is obtained by setting all vertices with
$m<2k$ to zero and performing the scaling $\beta-\beta_H\sim
g_s^{2(1-{1\ov k})}$, i.e. $a= 2 (1-{1 \ov k})$ in \fjsN.

Finally let us consider how to define various $\lam_{6}, \lam_8,
\cdots$ from string amplitudes. Recall that $\lam_4$ can be
obtained from \lamST\ and \jrjp. Naively one might want to define
$\lam_{2k}$ for $k=3, 4, \cdots$ by the {\it tree-level}
amplitudes of $k$ winding $1$ and $k$ winding $-1$ modes. However,
from factorization argument, these amplitudes are divergent at
$m_\phi^2 =0$. The divergences come from diagrams containing lower
order vertices $\lam_{2k'}$ with $k'<k$ and $\phi$ in the internal
propagators, which can be found from standard Feynman diagrams for
the action $m^2_\phi \phi \phi^* + \sum_{k'=2}^{k-1} \lam_{2k}
g_s^{2k-2}  (\phi \phi^*)^k$. $\lam_{2k}$ is thus given by the
sphere amplitude of $k$ winding $1$ and $k$ winding $-1$ modes
with the divergent parts subtracted.

\newsec{Hagedorn behavior from YM theories}

Our discussion in the last section was rather generic. In particular
it should apply to type IIB string theory in $AdS_5 \times S_5$ or
 other string theories in asymptotic AdS spacetime.
In an AdS spacetime with curvature radius $R$ much bigger than the
string and Planck lengths, there is a first order Hawking-Page
transition at temperature $T_{HP} \sim {1 \ov R}$ much below the
Hagedorn temperature $T_H \sim {1 \ov \sqrt{\apr}}$ at which the
thermal string gas in AdS becomes perturbatively
unstable~\refs{\Hawkingpage}. The discussion of the last section
describes what happens if one stays in the superheated thermal AdS
phase above the Hawking-Page temperature all the way to the
Hagedorn temperature. From the critical behavior at the Hagedorn
temperature one can then map out the potential for the thermal
scalar. Aspects of the Hagedorn transition in AdS have been
discussed in~\refs{\barbon,\lawrence}.

Hawking and Page's semi-classical discussion applies to IIB string
theory in AdS with a  cosmological constant small compared to the
string scale and to the Planck scale, which corresponds to $\NN=4$
super-Yang-Mills theory on $S^3$ at strong 't Hooft
coupling~\refs{\adscft}. At zero and weak 't Hooft coupling, which
is dual to a small AdS, thermodynamics of $\NN=4$ SYM theory on
$S^3$ has been discussed in~\refs{\sundB,\MinW}. In the free
theory limit the Hagedorn and Hawking-Page temperatures coincide.
At weak coupling it is not yet clear whether the transition is of
first or second order~\refs{\MinW}. Other studies of (Hagedorn)
phase transitions in Yang-Mills theories
include~\refs{\LiuVY,\AlvarezGaumeFV,\SchnitzerQT\SpradlinPP\AharonyBQ\GomezReinoBQ\BasuPJ
\FuruuchiST\AlvarezGaumeJG\BasuMQ\DeyDS\HikidaQB
\HarmarkTA\HarmarkET\SchnitzerXZ\AharonyBQ-\raamsdonk}.

In this section we show that the critical Hagedorn behavior found
in the last section arises also for a wide class of matrix quantum
mechanical systems including $\NN=4$ SYM on $S^3$. Our discussion
applies regardless of the order of the transition and also to
strong coupling. In particular, we show explicitly that the
Hagedorn divergences can be attributed to an effective potential
of the form \Mmbsdg, which was only argued in the last section.

The plan of this section is as follows. In next subsection we
introduce the family of theories to which our discussion applies,
which includes $\NN=4$ SYM on $S^3$. In the subsequent subsections
we discuss the large $N$ expansion of these theories at finite
temperature and identify new ingredients. We show that the free
energy contain contributions from ``vortices'' and introduce a set
of  vortex diagrams to describe them. The vortex diagrams can be
identified with degenerate worldsheets on the string theory side.
The critical behavior near the Hagedorn temperature and the
effective action for the thermal scalar are recovered at the end.

\subsec{Theories of interest}

Consider the following class of matrix quantum mechanical systems
 \eqn\onedAc{
 S =    \int_0^\beta\!d\tau  \;
\biggl[N \tr \sum_\al
 \le(\frac{1}{2} (D_\tau M_\al)^2
 - \ha \om_{\al}^2 M_\al^2 \ri) + N \tr \sum_{a}
 \xi_{a}^\dagger (D_\tau + \tilde \om_{a}) \xi_{a}
      + V (M_\al, \xi_a; \lam) \, \biggr]
 }
where:

\item{1.} We have written the action in Euclidean signature, with
the Euclidean time $\tau$ having a period $\beta = {1 \ov T}$. In
the zero temperature limit, $\beta \to \infty$.

\item{2.} $M_\al$ and $\xi_a$ are $N \times N$ bosonic and
fermionic matrices respectively, and
 \eqn\coDs{
D_\tau M_\al= \p_\tau M_\al - i [A, M_\al], \qquad
 D_\tau \xi_a= \p_\tau \xi_a - i [A, \xi_a] \ .
 }
are covariant derivatives. As a result, \onedAc\ has a $U(N)$ gauge
symmetry, with $M_\al, \xi_a$ transforming in the adjoint
representation. The  $(0+1)$-dimensional ``gauge field'' $A (\tau)$
plays the role of the Lagrange multiplier which imposes that
physical states are singlet of $U(N)$. $M_\al, \xi_a$ satisfy
periodic and anti-periodic boundary conditions respectively
 \eqn\ndbo{
 M_\al (\tau + \beta) = M_\al (\tau), \qquad \xi_a (\tau + \beta) = -\xi_a
 (\tau)\ .
 }

\item{3.} The frequencies $\om_\al$ and $\tilde \om_a$  in
\onedAc\ are nonzero for any $\al$ and $a$, i.e. the theory has a
mass gap and a unique vacuum. The number of matrices is greater
than one and can be infinite.

\item{4.} $V (M_\al, \xi_a;\lam)$ can be written as a sum of {\it
single-trace} operators and is controlled by a coupling constant
$\lam$, which remains fixed in the large $N$ limit.

\medskip

$\NN=4$ SYM on $S^3$ is an example of such systems with an
infinite number of matrices when the Yang-Mills and matter fields
are expanded in terms of spherical harmonics on $S^3$.
$V(M_\al,\xi_a;\lam)$ can be schematically written as\foot{The
precise form of the interactions depends on the choice of gauge.
It is convenient to choose Coulomb gauge $\nabla \cdot \vec A =0$,
in which the longitudinal component of the gauge field is set to
zero. In this gauge, $M_{\al}$ include also non-propagating modes
coming from harmonic modes of ghosts and the zero component of the
gauge field.}
 \eqn\higT{
 V =  N \le(\sqrt{\lam}  V_3 (M_\al,\xi_a) + \lam
 V_{4} (M_\al,\xi_a) \ri)
 }
 where $V_{3}$ and $V_{4}$ contain infinite sums of  single-trace operators
 which are cubic and quartic in $M_\al, \xi_a$. $\lam = g_{YM}^2 N$ is the 't~Hooft
 coupling.

To end this subsection, let us recall the standard relation
between the large $N$ expansion of  a matrix quantum mechanics
like \onedAc\ (or a gauge field theory) {\it at zero temperature}
with the string theory perturbative expansion~\refs{\thooft}. In
the large $N$ limit, the free energy of \onedAc\ can be organized
in terms of the topology of Feynman diagrams
 \eqn\ndjw{
 \log Z = \sum_{h=0}^\infty N^{2(1-h)} f_h (\lam)
 }
where $f_0 (\lam)$ is the sum of  connected planar Feynman
diagrams, and $f_1 (\lam)$ is the sum of  connected non-planar
diagrams which can be put on a torus, and so on.  The expansion
\ndjw\ resembles the perturbative expansion of a string theory,
with $1/N$ identified with the closed string coupling $g_s$ and
$f_h (\lam)$ identified with contributions from worldsheets of
genus-$h$. For $\NN=4$ SYM theory on $S^3$, $f_h (\lam)$ is the
contribution of string worldsheets of $h$ handles propagating in
$AdS_5 \times S_5$.

In the next few subsections, we discuss the large $N$ expansion of
\onedAc\ at finite temperature, and new ingredients arise. We find
new contributions associated with Feynman diagrams with vortices,
which can be identified with degenerate limits of a string
worldsheet. As a result, the same critical Hagedorn behavior is
recovered from gauge theories.

\subsec{Correlation functions in free theory}

In this subsection we discuss finite temperature correlation
functions of \onedAc\ in the free theory limit (i.e. with $V=0$),
focusing on the large $N$ counting. We will find that at finite
temperature, in addition to the standard $1/N^2$ corrections due
to non-planar diagrams, there are corrections due to
vortices. 
This subsection makes preparation for the discussion of the
interacting theory free energy in the next subsection.

\onedAc\ has a $U(N)$ gauge symmetry, which can be used to set the
gauge field $A (\tau)$ to zero. The gauge transformation, however,
modifies the boundary conditions from \ndbo\ to
 \eqn\nonTB{
 M_\al(\tau + \beta) = U M_\al U^\dagger ,
 \qquad \xi_a(\tau + \beta) = -U \xi_a U^\dagger \ .
 }
The unitary matrix $U$ can be understood as the Wilson line of $A$
wound around the $\tau$ direction (Polyakov loop), which cannot be
gauged away. Correlation functions can then be written in terms of a
path integral as
 \eqn\pathI{
 \vev{\cdots}_{0,\beta} = {1 \ov Z_0(\beta)} \int dU  \int D M_\al (\tau) D \xi_a(\tau) \,
 \cdots \; e^{-S_0 [M_\al, \xi_a; A =0]}
 }
with $M_\al, \xi_a$ satisfying the boundary conditions \nonTB. $S_0$
and $Z_0$ are the action and partition function for the free theory
respectively. The free theory action $S_0$ has only quadratic
dependence on $M_\al$ and $\xi_a$, thus the functional integrals
over these variables in \pathI\ can be carried out explicitly and
\pathI\ can be reduced to a matrix integral over $U$ only.

The free theory partition function can be written
as~\refs{\sundB,\MinW}
 \eqn\parT{
 Z_0 (\beta) = \int dU \; e^{I_{0} (U)}
 }
with $I_{0} (U)$ given by
 \eqn\wisp{
 I_{0} (U) = \sum_{n=1}^\infty {1 \ov n} V_n (\beta) \Tr U^n \Tr U^{-n}
 }
and
 \eqn\rnsp{
V_n (\beta) = z_b (n\beta) +
 (-1)^{n+1}
 z_f (n \beta) , \qquad z_b (\beta) = \sum_\al e^{-\beta \om_\al}, \qquad z_f
(\beta) = \sum_a e^{-\beta \tilde \om_a} \ .
 }
When the temperature $T$ is small, the matrix integral \parT\ can be
evaluated in the large $N$ limit as~\refs{\sundB,\MinW}
 \eqn\sbf{
Z_0 (\beta) =  C \prod_{n=1}^\infty {n \ov 1 - V_n (\beta)} +
 O(1/N^2)
 }
where $C$ is an $N$-independent constant factor. $Z_0 (\beta)$
becomes divergent if some $V_n (\beta)$ are equal to $1$. From
\rnsp\ one can check that $V_1 (\beta) > V_n (\beta)$ for $n
>1$ and that $V_1 (\beta)$ is a monotonically increasing function
of $T$, with $V_1 (\beta = \infty) =0$ and $ V_1 (\beta =0) > 1$.
Thus as one increases $T$ from zero, there exists a $T_H$, at
which $V_1 (T_H) = 1$ and $Z_0$ becomes divergent. Equation \sbf\
only applies to $T < T_H$. As pointed out in~\refs{\sundB,\MinW},
the divergence is precisely of the Hagedorn-type \fkd\ for a
string theory in a spacetime whose Laplacian has a gap. The
critical behavior of higher order terms in \sbf\ near $T_H$ and
the smoothing of the Hagedorn divergence at finite $N$ (i.e. in
quantum string theory) for free Yang-Mills theory was further
discussed in~\refs{\LiuVY}.

Correlation functions of  gauge invariant operators can be
obtained by first performing Wick contractions and then evaluating
the matrix integral for $U$. With boundary conditions \nonTB, the
contractions of $M_a$ and $\xi_a$ are~\refs{\brigante}
 \eqn\fincon{\eqalign{
 & \underbrace{M_{ij}^a (\tau) \, M_{kl}^b (0)}
 = {\delta_{ab} \ov N} \sum_{m=-\infty}^\infty G_s(\tau - m \beta; \om_a)
 U^{-m}_{il} U^{m}_{kj}
 \cr
 & \underbrace{\xi_{ij}^a (\tau) \, \xi_{kl}^b (0)}
 = {\delta_{ab} \ov N} \sum_{m=-\infty}^\infty (-1)^m G_f(\tau - m \beta; \tilde \om_a)
 U^{-m}_{il} U^{m}_{kj} \
 \cr
 }}
where $G_s$ and $G_f$ are standard $(0+1)$-dimensional propagators
at zero temperature
 \eqn\oenrP{ G_s (\tau; \om)  = {1 \ov 2
\om} e^{-\om|\tau|}, \qquad
 G_f(\tau; \om)  = (-\d_\tau +\om
)G_s(\tau;\om) \ .
 }

\ifig\finTDia{An example of a double-line diagram at finite
temperature. Each propagator carries a winding number (or image
number), which should be summed over. Due to the presence of
$U$-factors in \fincon, associated with each face one finds a
factor of $\tr U^{s_A}$, instead of a factor $N$ as is the case at
zero temperature.} {\epsfxsize=4cm \epsfbox{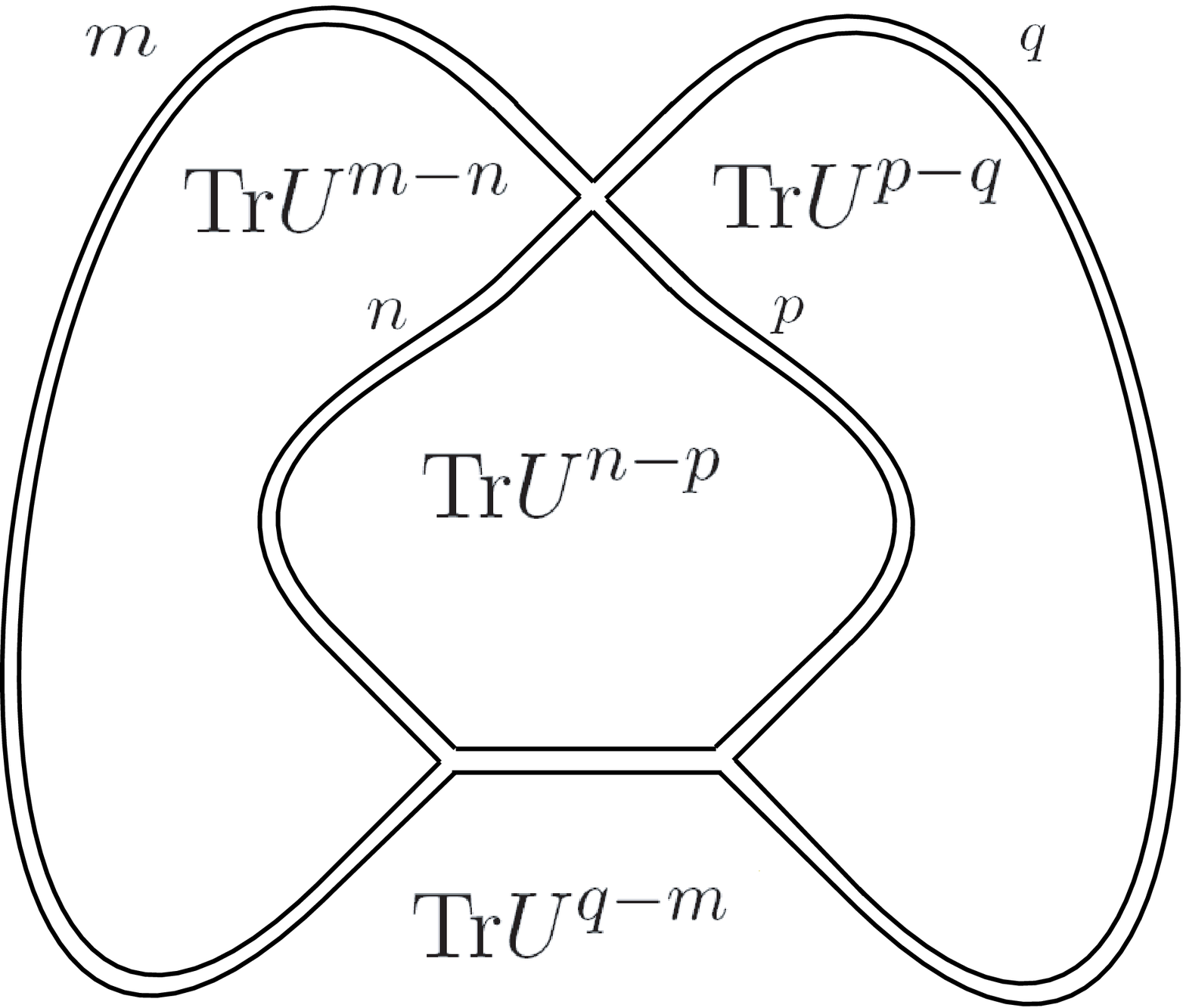}}

It follows from \fincon\ that at finite temperature, each
propagator in a double-line Feynman diagram carries a winding
number (or image number) $m$, which should be summed over (see
\finTDia). More explicitly, using \fincon, the contribution of a
generic Feynman diagram to a correlation function of single-trace
operators\foot{We assume the operators are normalized as $N \tr
(\cdots)$.} can be written in the form\foot{For notational
simplicity, we only include bosonic modes in the equation below.
It can be easily generalized to include
fermions.}~\refs{\brigante}
 \eqn\genrTm{
  {1 \ov N^{2h-2}} \le(\prod_{i \leq j} \prod_{p=1}^{I_{ij}}
  \sum_{m_{ij}^{(p)}=-\infty}^\infty \ri)
  \le(\prod_{i \leq j} \prod_{p=1}^{I_{ij}}
 G_s^{(p)} \le(\tau_{ij} - m_{ij}^{(p)} \beta\ri) \ri) 
 \vev{{1 \ov N} \tr U^{s_1} {1 \ov N} \tr U^{s_2} \cdots {1 \ov N} \tr U^{s_F}}_U
 }
where $i,j$ enumerate the vertices (i.e. operator insertions) and
$p$ enumerates the propagators between vertices $i$ and $j$ with
$I_{ij}$ the total number of propagators between them.
$m_{ij}^{(p)}$ label the images of $G^{(p)} (\tau_{ij})$. $h$ is
the genus of the diagram. In \genrTm,
 \eqn\usna{
 \vev{\cdots}_U = {1 \ov Z_0 (\beta)} \int dU \; \cdots \;
e^{I_{0} (U)}
 }
with $I_{0} (U)$ given by \wisp. The powers $s_1, s_2, \cdots$ in
the last factor of \genrTm\ can be found as follows. To each
propagator in the diagram we assign a direction and an orientation
can be chosen for each face. For each face $A$ in the diagram, we
have a factor $\tr U^{s_{A}}$, with $s_A$ given by
 \eqn\idnd{
 s_A = \sum_{\p A} (\pm) m_{ij}^{(p)}, \qquad A =1,2, \cdots F
 }
where the sum $\p A$ is over the propagators bounding the face $A$
and $F$ denotes the total number of faces in a diagram (see e.g.
\finTDia). In \idnd\ the plus (minus) sign is taken if the
direction of the corresponding propagator is the same as (opposite
to) that of the face. $s_A$ has a precise mathematical meaning: it
is the number of times that the Euclidean time circle is wrapped
around by the propagators bounding a face $A$. We will thus call
$s_A$ the vortex number for face $A$. Note that since the
exponents of $U$ add up to zero in \fincon, for each {\it
connected part} of a Feynman diagram the sum of $s_A$ adds to
zero. To illustrate more explicitly how \genrTm\ works, we give
some examples in Appendix A.

The partition function \parT\ and more generally matrix integrals in
\genrTm\ can be evaluated to all orders in a $1/N^2$ expansion. In
Appendix B we prove that, {\it up to corrections non-perturbative in
$N$}, the matrix integrals can be evaluated by treating each $\Tr
U^n$ as an independent integration variable. More explicitly, \usna\
can be evaluated by replacing
 \eqn\fjje{
  {1 \ov N} \Tr U^n \to \phi_n , \qquad {1 \ov N} \Tr U^{-n} \to
  \phi_{-n} =
  \phi_n^*, \qquad \phi_0 =1 \ ,
 }
i.e.
 \eqn\corrTrU{\eqalign{
  & \vev{{1 \ov N} \Tr U^{s_1} {1 \ov N} \Tr U^{s_2} \cdots {1 \ov N} \Tr U^{s_F}}_U \cr
  & = {1 \ov Z_0} \int_{-\infty}^\infty \le(\prod_{i=1}^\infty d \phi_i d
  \phi_i^*\ri)
  \; \phi_{s_1} \cdots \phi_{s_F} \; \exp \le(- N^2 \sum_{n=1}^\infty
  {v_n (\beta)} \phi_n \phi_n^* \ri) \cr
   & \qquad  + \quad {\rm nonperturbative \;\; in} \; N
  \cr
 }}
where
  \eqn\rjsj{
v_n (\beta) = {1 - V_n (\beta) \ov n} \ .
 }
From \corrTrU,
 \eqn\rTrU{\eqalign{
  & \vev{{1 \ov N} \Tr U^{s_1} {1 \ov N} \Tr U^{s_2} \cdots {1 \ov N} \Tr U^{s_F}}_U \cr
 & = \prod_{i=1}^{F}  \delta_{s_i,0} +{1 \ov N^2}
 \sum_{i<j=1}^{F} \le( {1 \ov v_{|s_i|} (\beta)} \, \delta_{s_i + s_j,0}
\prod_{k=1\;k \neq i, j}^{F}  \delta_{s_k,0} \ri) \cr
 & \qquad + O(N^{-4}) + {\rm nonperturbative \;\; in} \; N
 }}
where order $1/N^2$ terms are obtained by contractions of one pair
of $\phi_{s_i}$'s, order $1/N^4$ terms are obtained by contracting
two pairs of $\phi$'s, and so forth. Each contraction brings a
factor of ${1 \ov N^2 v_{s_i} (\beta)}$. Perturbative corrections in
$1/N^2$ terminate at order $1/N^{F}$ (or $1/N^{F-1}$) for $F$ even
(odd). For example, there is no other perturbative correction in
$1/N^2$ for the partition function \sbf, and for $F=2$
 \eqn\tjs{\eqalign{ \vev{{1 \ov N} \tr U^n {1 \ov N} \tr
U^{m}} & =  \delta_{n,0} \delta_{m,0}+ {1 \ov N^2} {1 \ov v_{|n|}
(\beta)} \delta_{m+n,0} + {\rm nonperturbative \;\; corrections} \
.
 }}

To summarize, combining \genrTm\ and \rTrU\ we find that for a
correlation function of gauge invariant operators, there are two
sources of $1/N^2$ corrections:

\item{1.} From the genus of the diagram as indicated by the power
of $1/N$ in \genrTm. This follows from the standard large $N$
counting.

\item{2.} From the $1/N^2$ corrections of the matrix integral
\rTrU. The leading order term in \rTrU\ imposes the constraint that
for any face $A$ of the diagram the vortex number $s_A$ should be
zero. The next order corresponds to having nonzero vortex numbers in
two of the faces, say the faces $A$ and $B$ with $ s_A s_B \neq 0$
and $s_A+ s_B=0$. Below, we will refer to those diagrams with
nonzero vortex numbers as containing vortices, in anticipation of
their interpretation from the string worldsheet\foot{See also the
discussion of~\refs{\klebanov} in the context of $c=1$ matrix
models.}. From remarks below~\idnd, if a face A of a Feynman diagram
contains a vortex with vortex number $s_A$, then the propagators
bounding the face wrap around the Euclidean time circle $s_A$ times.
At finite temperature, due to the presence of vortices, planar
diagrams also contain higher order $1/N^2$ corrections.

\ifig\vorteX{Examples of double-line diagrams with nonzero
vortices. Each thin line~(vortex propagator) represents a
contraction in \rTrU. Compare the left diagram to \finTDia.
Diagrams which are disconnected at zero temperature can be
connected through vortex propagators as in the right diagram.}
{\epsfxsize=7cm \epsfbox{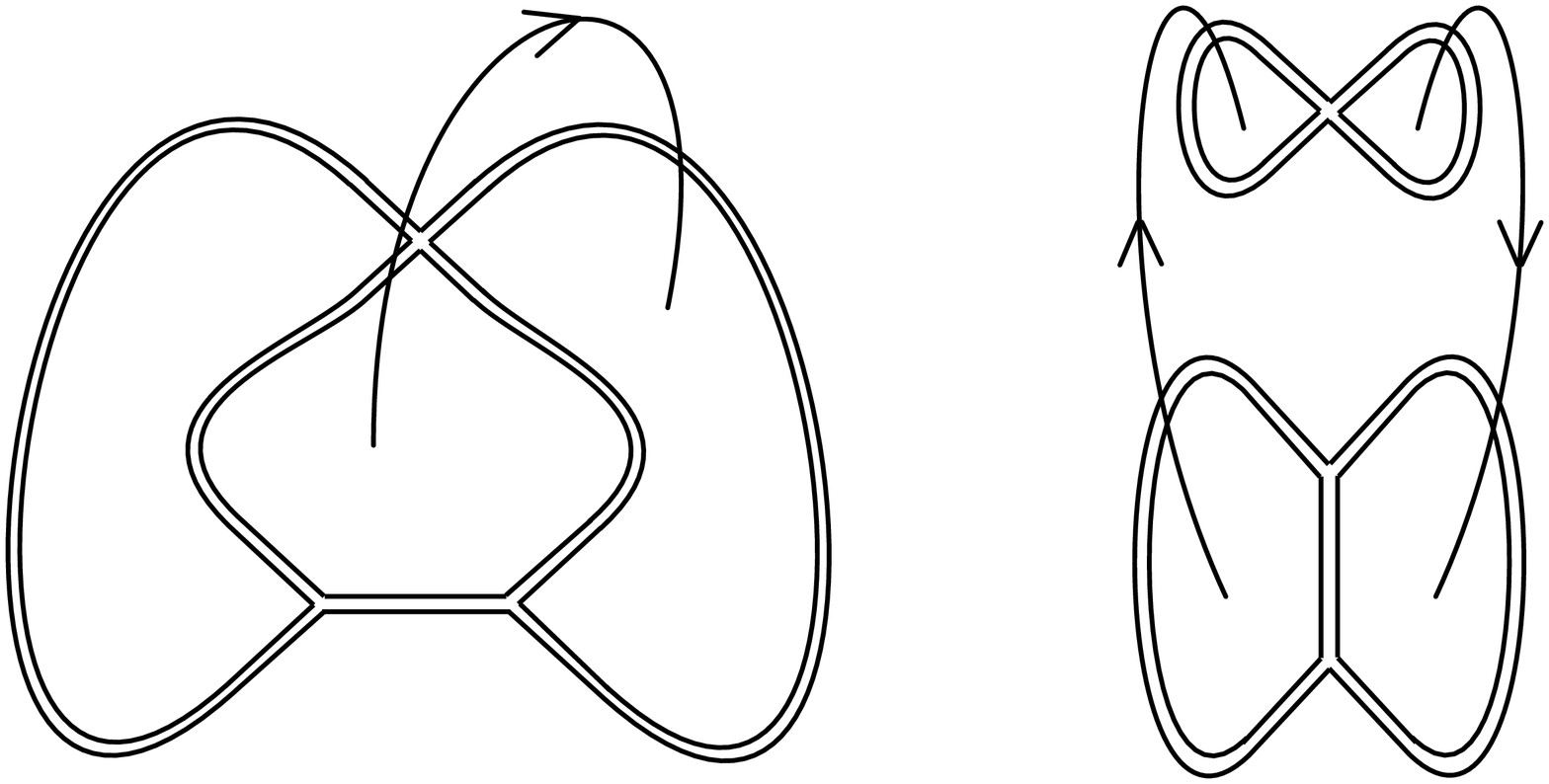}}

\ndt  It will be convenient to represents the vortex contributions
diagrammatically: we represent each contraction in \rTrU\ by an
oriented line between two surfaces which have the opposite vortex
numbers. The orientation of a line is that it exists (enters) the
surface if its vortex number is positive (negative). We associate a
factor $1/N$ for each vortex and a factor $1/v_n (\beta)$ to a line
(vortex propagator) connecting two surfaces with vortex number $\pm
n$. See \vorteX\ for some examples of such diagrams. Note that a
diagram with otherwise disconnected parts connected by vortex lines
should be considered as connected, as in the right diagram of
\vorteX. In computing a correlation function one should sum over all
possible vortex contributions.

To summarize this subsection, in computing correlation functions at
finite temperature, one should consider not only Feynman diagrams
which appear at zero temperature, but also diagrams with nonzero
vortices. Explicit examples are given in Appendix~A.

\subsec{Free energy in interacting theory and vortex diagrams}

We now consider the Euclidean partition function of the interacting
theory. Our purpose is to identify $T_H$ and the critical behavior
near $T_H$ to all orders in the $1/N^2$ expansion.

In perturbation theory, the partition function can be evaluated by
expanding the interaction terms in the exponent of the path
integral
 \eqn\nshj{
 Z (\beta, \lam) = Z_0 (\beta) \sum_{n=0}^\infty {(-1)^n \ov n!} \int_0^\beta \prod_{i=1}^n d
 \tau_i \, \vev{V(\tau_1) \cdots V (\tau_n)}_{0, \beta}
 }
In \nshj, $\vev{\cdots}_{0,\beta}$ denotes free theory correlation
functions and recall that $V$ is given by a sum of single trace
operators of the form $N \tr (\cdots)$. The free energy can be
obtained from
 \eqn\rjsZ{
 \log Z = \log Z_0 + \sum_{n=0}^\infty {(-1)^n \ov n!} \int_0^\beta \prod_{i=1}^n d
 \tau_i \, \vev{V(\tau_1) \cdots V (\tau_n)}_{0, \beta, connected}
 }
i.e. one sums only over the connected diagrams. The discussion in
the last subsection for free theory correlation functions can now be
directly carried over to $\log Z$. In particular, there are two
sources of $1/N^2$ corrections: from the non-planar structure and
from vortices. We can expand $\log Z$ in $1/N^2$ as
 \eqn\rjee{
\log Z (\beta) = \sum_{n=0}^\infty N^{2-2n} \ZZ_n (\beta)
 = N^2 \ZZ_0 (\beta)
 + \ZZ_1 (\beta) + {1 \ov N^2} \ZZ_2 (\beta) + \cdots
 }
where $\ZZ_{0}$ corresponds to the sum over connected planar
diagrams with no vortices, while $\ZZ_1$ contains the sum  of
connected genus-$1$ non-planar diagrams with no vortices {\it and}
planar diagrams with one pair of vortices, and so on. Recall that
each vortex carries a factor $1/N$ and they always come in pairs.
Also as remarked at the end of the last subsection, a diagram with
otherwise disconnected parts connected by vortex propagators is
connected.

\ifig\Vdiag{The propagators and vertices for vortex diagrams. The
vertices $Q^{(h,n)}$ of a vortex diagram have $n$ legs, each of
which is labelled by a vortex number. The sign of the vortex
number is positive (negative) if the corresponding leg exists
(enters) the vertex. The total vortex number of a vertex is zero.
We show $Q^{(0,2)}$, $Q^{(1,3)}$ in the figure as illustrations.}
{\epsfxsize=12cm \epsfbox{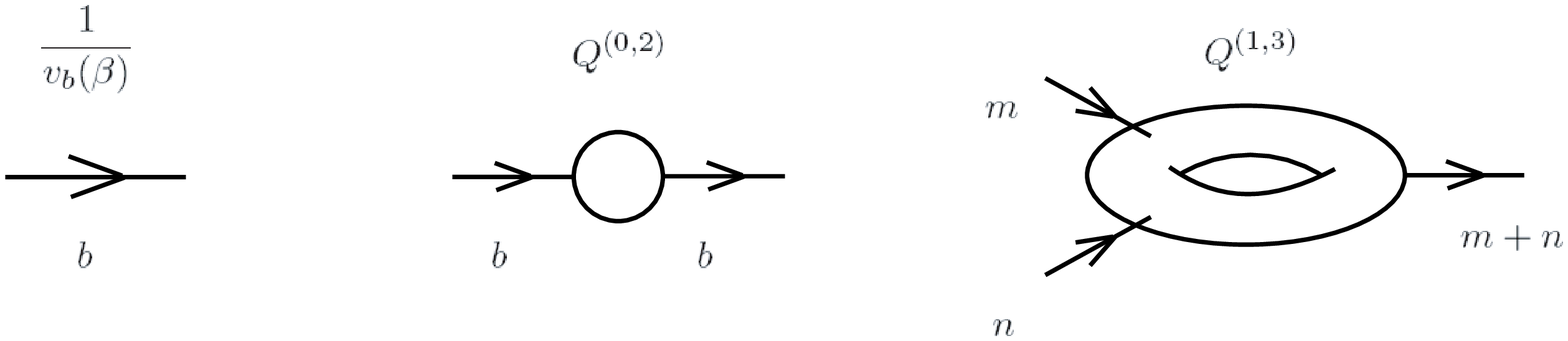}}

To elucidate the structure of  $\ZZ_g$, we introduce a new set of
``vortex diagrams'', by generalizing the diagrammatical rules
introduced below \vorteX:

\item{1.}  Denote $Q^{(h,n)}$ as the sum of
connected 
Feynman diagrams with genus $h$ and with $n$ vortices. In terms of
large $N$ counting, $Q^{(h,n)}$ is of order $N^{2- 2 h - n}$, as we
associate a factor $1/N$ with each vortex. Each vortex is labeled by
a vortex number and the total vortex number carried by $Q^{(h,n)}$
is zero\foot{ This follows from the discussion below \idnd.}.
Diagrammatically, $Q^{(h,n)}$ are represented as vertices with $n$
oriented legs. The leg exits the vertex if the corresponding vortex
number is positive.

\item{2.} Vortex diagrams are then constructed following the usual
rules with $Q^{(h,n)}$ as fundamental vertices and $1/v_b (\beta),
b > 0$ as propagators. Note that $b$ is the  vortex number carried
by a propagator and $v_b$ was defined in \rjsj.

\item{3.} The combinatoric rules are the same as standard Feynman
diagram. In particular, if there are $m$ identical vertices
$Q^{(h,n)}$ in a diagram, there is a factor $1/m!$, which comes
from the fact that disconnected diagrams are obtained from
connected ones by exponentiation.

\ndt Using the above diagrammatical rules, we now enumerate the
contributions to~$\ZZ_g$. See \Vdiag\ for illustrations of
propagators and vertices for vortex diagrams.

Let us first look at $\ZZ_0$, which is given by the sum of all
planar diagrams without vortex. In section 4 of~\refs{\brigante} it
was shown that $\ZZ_0$ is identical to the corresponding expression
at zero temperature and thus is temperature-independent\foot{$\ZZ_0$
is a special case of the discussion in section 4 of~\refs{\brigante}
with no external operator insertions.}. Since the free energy
$-\beta F$ is defined by subtracting the zero-temperature
contribution (which is the vacuum energy) from \rjee, we conclude
that the planar contribution to the free energy is identically
zero\foot{as is the case for a string theory below the Hagedorn
temperature.}.

\ifig\nonPla{Vortex diagrams contributing to $\ZZ_1^{(3)}$.}
{\epsfxsize=10cm \epsfbox{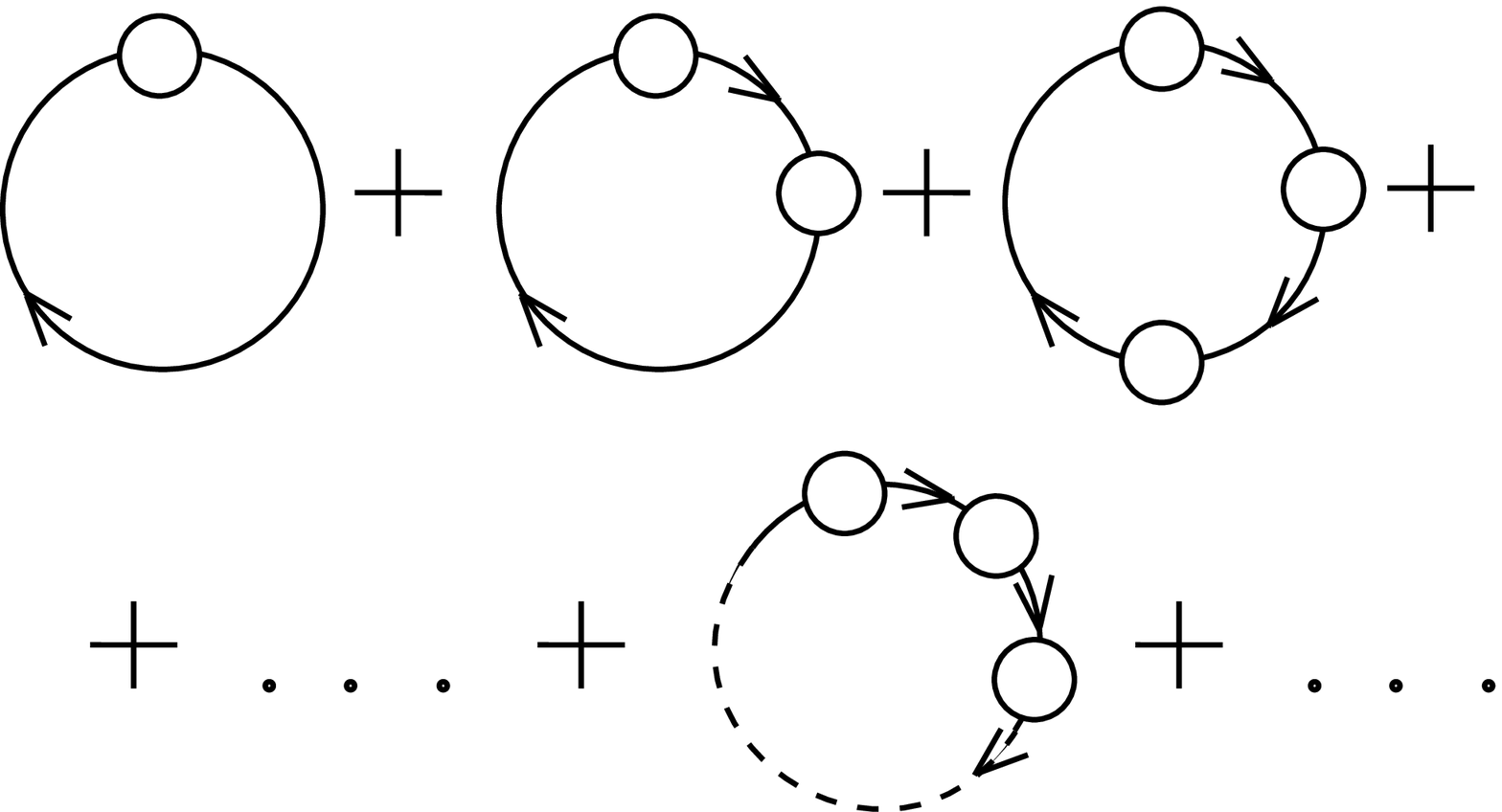}}

We now look at $\ZZ_1$, which contains three contributions: (i)
genus-1 contribution in free theory coming from the first term in
\rjsZ; (ii) sum of genus-1 Feynman diagrams with no vortices; (iii)
planar diagrams with vortices. The first contribution $\ZZ_1^{(1)}$
is given by the logarithm of \sbf. The second contribution
$\ZZ_1^{(2)}$ is given by $Q^{(1,0)}$. To find the third
contribution $\ZZ_1^{(3)}$, let us denote $Q^{(0,2)}_b$ the sum of
all planar connected diagrams with two vortices of winding $\pm b$.
Graphically, it can be represented by a sphere with an arrow
pointing in and an arrow pointing out, each carrying vortex number
$b$, as in the second diagram of \Vdiag. Using $Q^{(0,2)}_b$,
$\ZZ_1^{(3)}$ is obtained  by summing the vortex diagrams in
\nonPla. The combinatoric factor for a diagram with $m$ vertices is
$1/m$ following from the cyclic symmetry and we find that
 \eqn\logZinser{
 \ZZ_1^{(3)} =  \sum
_{b=1} ^\infty \sum _{m=1} ^\infty {1\ov m}
\left({Q^{(0,2)}_b(\lambda, \beta)\ov v_b (\beta)}\right) ^m =
 -\sum _{b=1}^\infty \log\left(1- {Q^{(0,2)}_b(\lambda,
\beta)\ov v_b (\beta)}\right)
 \ .
 }
Adding all three contributions together we find that
 \eqn\Zsaddle{\eqalign{
 \ZZ_1 & = \ZZ_1^{(1)} + \ZZ_1^{(2)} + \ZZ_1^{(3)} \cr
 & = Q^{1,0} (\beta, \lam)
 -\sum _{b=1}^\infty \left(\log\left(1-
{Q^{(0,2)}_b(\lambda, \beta)\ov v_b (\beta)}\right) + \log v_b
(\beta) \right) \cr
 & = Q^{1,0} (\beta, \lam) -\sum _{b=1}^\infty \log\left(v_b (\beta) -
{Q^{(0,2)}_b(\lambda, \beta)}\right) \ .
 }}

\ifig\ProaG{The dark thick line represents the resumed propagator
$\GG_b = {1 \ov v_b - Q^{(0,2)}_b}$.} {\epsfxsize=10cm
\epsfbox{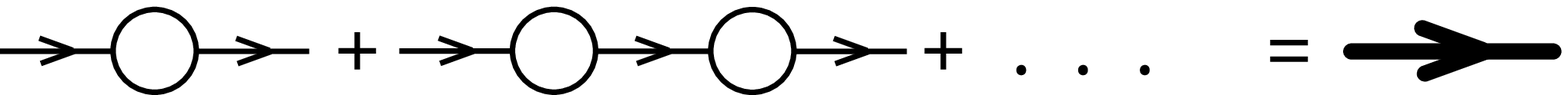}}

It should be clear from the above discussion of $\ZZ_1^{(3)}$ that
$Q^{(0,2)}_b$ should not really be treated as a vertex. Rather all
$Q^{(0,2)}_b$ should be resumed along with the propagators ${1 \ov
v_b (\beta)}$ to obtain a "resumed propagator" for each vortex
number
 \eqn\summprop{
 \GG_b (\beta) = \sum_{n=1}^\infty {1 \ov v_b^n (\beta)} (Q^{(0,2)}_b)^n
 = {1\ov v_b -Q^{(0,2)}_b}
 }
as shown diagrammatically in \ProaG. Note that \Zsaddle\ can be
rewritten in terms of $\GG_b$ as
 \eqn\rusn{
 \ZZ_1 = Q^{1,0} (\beta, \lam) + \sum _{b=1}^\infty \log \GG_b (\beta) \ .
 }

\ifig\exmD{Vortex diagrams contributing to $\ZZ_2$. Compare plots
in \exmD\ with the degenerate limits of a genus-$2$ surface in
\genTwo. Note the 2nd, 3rd and 5th diagrams in \genTwo\ do not
 appear in the above since they contain propagators which have to have
 zero windings.}
{\epsfxsize=10cm \epsfbox{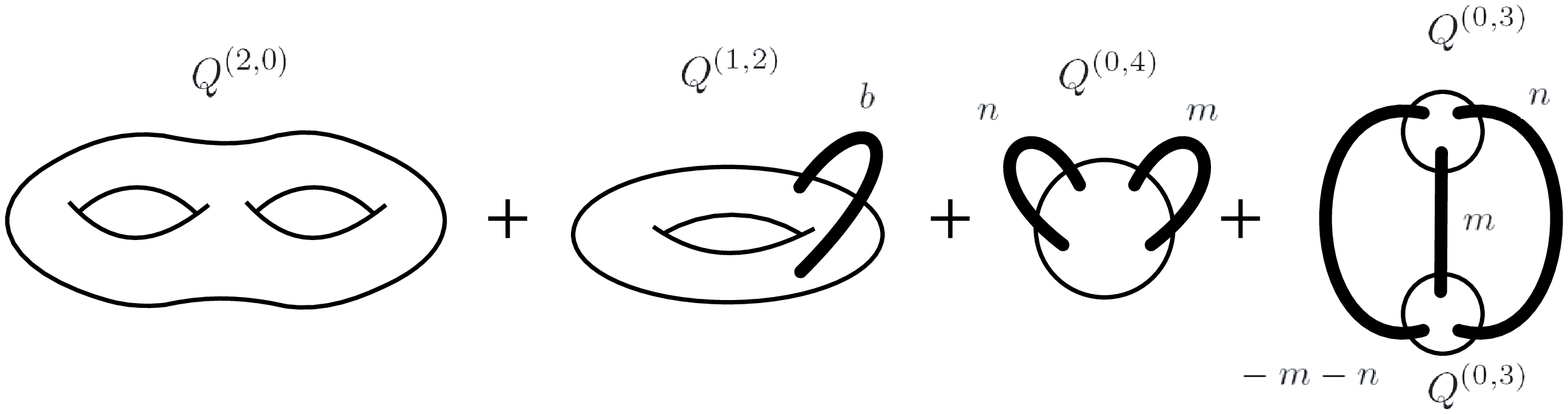}}

In the vortex diagrams for $\ZZ_g$ with $g \geq 2$, only resumed
propagators $\GG_b$ appear. As an example, the vortex diagrams
contributing to $\ZZ_2$ are shown in \exmD. Higher order diagrams
contributing to general $\ZZ_g$ can be similarly constructed.

By now readers may have recognized the resemblance of vortex
diagrams with the diagrams in \Degen\ and \genTwo. Indeed it is
natural to identify  vortex diagram contributions in the gauge
theory with contributions from degenerate limits of string
worldsheets in the corresponding string theory. For example,
diagrams in \exmD\ can be identified with various degenerate limits
(\genTwo) of genus two Riemann surfaces. In particular, vortices in
gauge theory vortex diagrams can be identified with insertions of
winding tachyon modes in the worldsheet. On the worldsheet if one
follows a closed contour around the vertex operator of a winding
tachyon mode of winding number $b$, the Euclidean time circle is
traversed $b$ times. Similarly, as discussed earlier if a face of a
Feynman diagram contains a vortex with vortex number $b$, the
propagators bounding the face wrap around the Euclidean time circle
$b$ times.

A more careful comparison between vortex diagrams for $\ZZ_g$ and
degenerate limits of a genus-$g$ surface (e.g. between \exmD\ and
\genTwo) also show some important differences:

\item{1.}  Notice that the 2nd, 3rd and 5th diagrams in \genTwo\
do not appear in \exmD. These diagrams are distinguished in that
some
 propagators are forced to have zero winding due to winding number conservation.
One can convince oneself that this feature persists to all orders.
Thus YM vortex diagrams do not correspond to the full
contributions from degenerate limits of a Riemann surface. All
propagators in the YM vortex diagrams carry nonzero windings.

\item{2.}Various degenerate limits of a Riemann surface do not
follow the standard Feynman rules and cannot be treated as Feynman
diagrams. For example, the third diagram of \genTwo\ can be obtained
as a degenerate limit of the first diagram and the fifth as a limit
of the fourth, etc. In contrast, the  vortex diagrams we constructed
in Yang-Mills theory do follow standard Feynman rules. In
particular, different diagrams in \exmD\ do not overlap.

\ndt Thus vortex diagrams correspond to a specific decomposition
of the boundary of the moduli space and can be considered as
defining an effective string field theory for the winding tachyon
modes.

\subsec{Critical Hagedorn behavior and the effective action}

Now let us examine the critical Hagedorn behavior of \rjee\ by
increasing the temperature from zero.

In free theory, as reviewed after equation \sbf, there is a Hagedorn
temperature given by equation $V_1 (\beta_H) =1$ at which the free
energy diverges as $\log Z_0 \approx - \log (\beta - \beta_H)$. Note
that there is {\it only} a one-loop divergence since all
perturbative corrections in $1/N$ to \sbf\ vanish.

In the interacting theory the effective vertices $Q^{(h,n)}$
should be regular at any temperature since they involve only sums
of products of \oenrP\ and their images. The divergences of
$\ZZ_n$ then can only occur when the resumed propagator $\GG_b
(\beta)$ \summprop\ become divergent, which happens when
 \eqn\sbjr{
 v_b (\beta) = Q_b^{(0,2)} (\lam,\beta) , \qquad {\rm i.e.} \qquad
  {1 - V_b (\beta) \ov b} = Q_b^{(0,2)} (\lam, \beta), \qquad b
  =1,2 \cdots .
  }
If we again assume that \sbjr\ is first satisfied for $b=1$ as one
decreases $\beta$ from infinity\foot{which should be the case for
$\lam$ small since $Q_b^{(0,2)}$ starts at order $O(\lam)$. For
large $\lam$, in principle this does not appear to be guaranteed
from the gauge theory point of view. However, from string theory it
appears always to be  the case that the lowest winding modes become
massless first.}, the Hagedorn temperature in the interacting theory
is determined by
 \eqn\sjjs{
 V_1 (\beta_H (\lam)) = 1 - Q_1^{(0,2)} (\lam,  \beta_H (\lam))
 }
with the most divergent term in $\ZZ_1$ given by (see~\Zsaddle)
 \eqn\hbsa{
 \ZZ_1 \approx - \log (\beta - \beta_H (\lam)) + {\rm finite}, \qquad
 \beta \sim \beta_H (\lam) \ .
 }
Divergences in $\ZZ_n$ can be analyzed following exactly the same
power counting argument of the last section (after equation~\rjj).
We find that the most divergent contribution to $\ZZ_n$ as $\beta
\to \beta_H$ is given by
 \eqn\jeue{
 {1 \ov (\beta - \beta_H)^{2n}} \ .
 }
Furthermore, since the construction of vortex diagrams follows the
standard combinatoric rules of Feynman diagrams, we find that the
most divergent pieces at each $1/N^{2h}$ order is precisely given by
\djsa\ with the identification
 \eqn\jejn{
  m^2_{\phi} = v_1 (\beta) - Q_1^{(0,2)} (\lam,\beta), \qquad
 {\lam_4 \ov N^2} = Q^{(0,4)}_{1,1, -1,-1} + {1\ov v_2 (\beta)-
Q^{(0,2)}_2} Q^{(0,3)}_{1,1,-2} Q^{(0,3)}_{-1,-1,2}
 }
where the subscripts in $Q^{(h,n)}$ denote the vortex numbers for
each leg. Similarly, one can use the same argument before \Mmbsdg\
to  extract higher order terms in the effective action \Mmbsdg.
The fact that we get \Mmbsdg\ from divergences is guaranteed since
vortex diagrams follow the rules of Feynman diagrams. It is also
straightforward to work out the counterparts of \jejn\ between
$\lam_{2k}$ in string theory side and $Q^{(m,n)}$.

We note that on general grounds one expects that the free energy of
the interacting theory can be written in terms of a matrix integral
for $U$~\refs{\MinW}
  \eqn\parTI{
 Z (\beta,\lam) = \int dU \; e^{I (U)}
 }
with $I (U)$ expanded in terms of all possible powers of $\tr U^n$
 \eqn\wisp{\eqalign{
 I (U) & = Q^{(0)} +
\sum_{n\neq 0}  Q^{(2)}_n \tr U^n \tr U^{-n} + \sum_{^{nml\neq
0}_{n+m+l=0}} Q_{nml}^{(3)} \tr U^n \tr U^m \tr U^{l} \cr
 + & \sum_{^{n,m,l,p\neq0}_{n+m+l+p
=0}}Q_{nmlp}^{(4)} \tr U^n \tr U^m \tr U^l \tr U^{p} + \cdots
 }
 }
where each $Q^{(n)}_{\cdots} = \sum_{h=0}^\infty
Q^{(h,n)}_{\cdots}$ is a sum of contributions of diagrams of
different genus $h$ and $\cdots$ denotes windings of insertions.
The vortex diagrams introduced earlier can be considered as the
diagrammatical rules for computing
\parTI. The effective action~\Mmbsdg\ then extracts the most
important contribution near the Hagedorn temperature.

It is important to emphasize that our discussion above should also
apply to strong coupling. $Q^{(n,m)} (\lam)$, which are the basic
building blocks of the vortex diagrams, can be defined
non-perturbatively as follows. Since at each genus the number of
Feynman diagrams grows with loops only as a power, we expect that
$Q^{(n,m)} (\lam)$ should have a finite radius of convergence in the
complex $\lam$ plane. Once one obtains $Q^{(n,m)} (\lam)$ near the
origin, one can analytically continue them to strong coupling.

\newsec{Conclusions and discussions}

In this paper we extracted Hagedorn divergences to all string loop
orders and showed that they can be resumed. The resumed amplitudes
have the form of an integral over the potential \ejro\ for the
thermal scalar and smooth the divergences. We presented arguments
both from a worldsheet approach and from Yang-Mills theories using
AdS/CFT. In the double scaling limits \fjsN, worldsheets with
arbitrary number of thermal scalar insertions become equally
important, which is consistent with the expectation that the
thermal scalar will condense and the spacetime background will
shift.

The fact that one can obtain the thermal scalar potential to
arbitrary higher orders by analyzing the local divergences in the
thermal string phase is interesting. The potential would enable
one to find other possible phases of the theory. The results also
give an unambiguous prescription for computing the potential for
the thermal scalar near the Hagedorn temperature from string
amplitudes. The relation we found between vortex diagrams in
Yang-Mills theory at finite temperature and degenerate limits of
worldsheet Riemann surfaces is rather intriguing and worth
investigating further.

Finally we note our strategy for extracting the thermal scalar
potential should also be applicable to the tachyon condensation in
a circle with anti-periodic boundary conditions~(for a recent
discussion see~\refs{\dineetal}).

\bigskip
\noindent{\bf Acknowledgments}

We would like to thank I.~Klebanov, A.~Lawrence, J.~McGreevy,
J.~Polchinski, S.~Shenker, S.~Wadia and B.~Zwiebach for
discussions and P.~Talavera for collaboration at early stage of
this work. This work is supported in part by Alfred~P.~Sloan
Foundation, U.S. Department of Energy (D.O.E) OJI grant, funds
provided by the U.S. Department of Energy (D.O.E) under
cooperative research agreement \#DF-FC02-94ER40818, and in part by
National Science Foundation under Grant No. PHY05-51164.

\appendix{A}{Examples of \genrTm}

 \ifig\PlanExpD{Planar disconnected contributions to
$\vev{\Tr M^4(\tau) \Tr M^4(0)}$}{\epsfxsize=6cm
\epsfbox{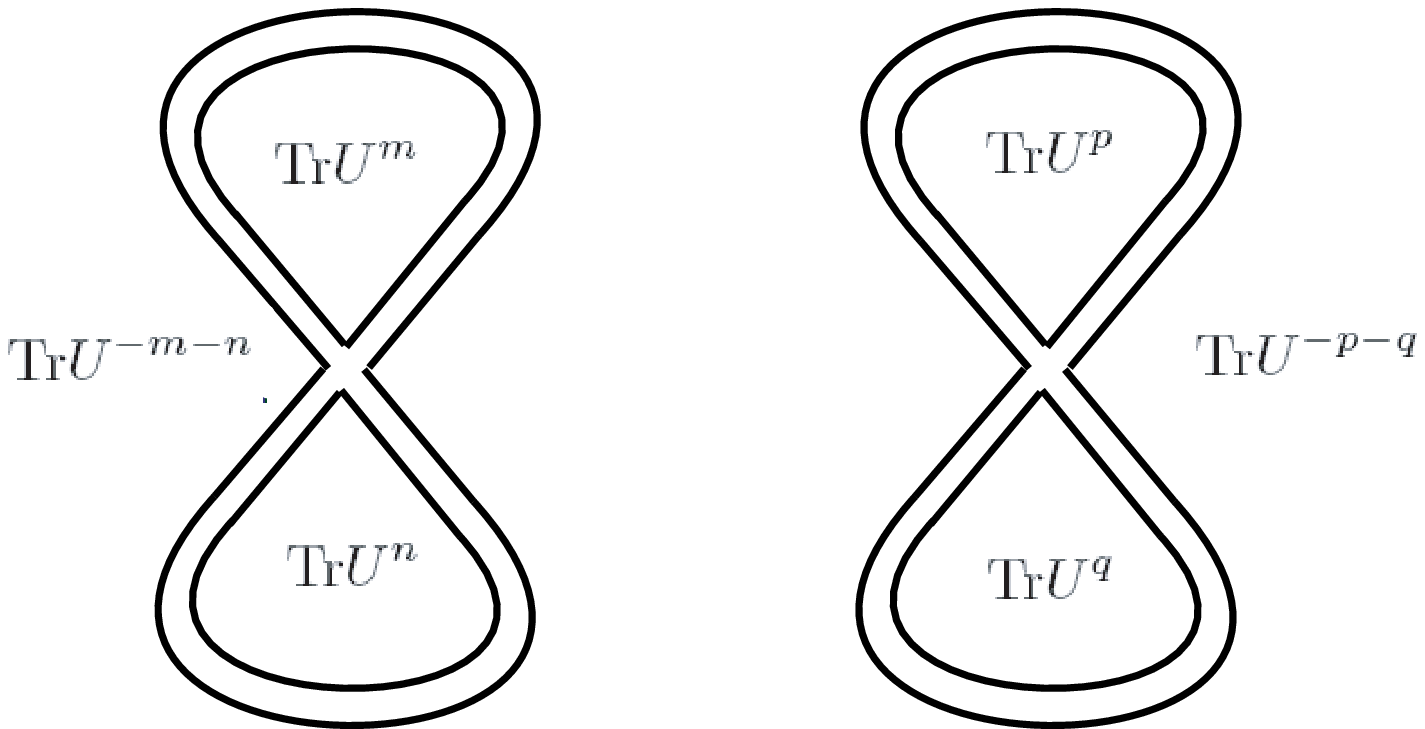} }

 In this appendix we give some explicit
examples on the use of equation \genrTm\ for calculating
correlation functions between single trace operators. For
definiteness we consider only bosonic operators, but the procedure
is analogous for operators involving fermions. Consider the
following simple example
 \eqn\trmftrmf{
  \vev{N \Tr M^4(\tau)
 N \Tr M^4(0)}}
where $M$ can be any of the bosonic modes in \onedAc. The
calculation of \trmftrmf\ amounts to drawing all possible double
line diagrams. For example the disconnected planar contribution is
given in \PlanExpD. From \fincon, each propagator carries an image
number (or winding number), which should be summed over. Each face
$A$ carries a factor $\tr U^{s_A}$. $s_A$ is determined by
choosing a direction for the propagators, and an orientation for
the face, as explained below \idnd. \PlanExpD\ therefore gives a
contribution of the form
  \eqn\contone{\eqalign{
 {4\ov N^2}\!\! \!\!\!\! \sum_{m,n,p,q=-\infty}^\infty \!\! \!
 \!\!\!\!
 G_s(- m \beta)G_s(- n
\beta)G_s(- p \beta)G_s(- q \beta) 
\vev{\Tr U^m \Tr
U^n \Tr U^{-m-n} \Tr U^p \Tr U^q \Tr U^{-p-q}}_U}  \ .}

\ifig\PlanExp{Planar connected contributions to $\vev{\Tr
M^4(\tau) \Tr M^4(0)}$}{\epsfxsize=12cm \epsfbox{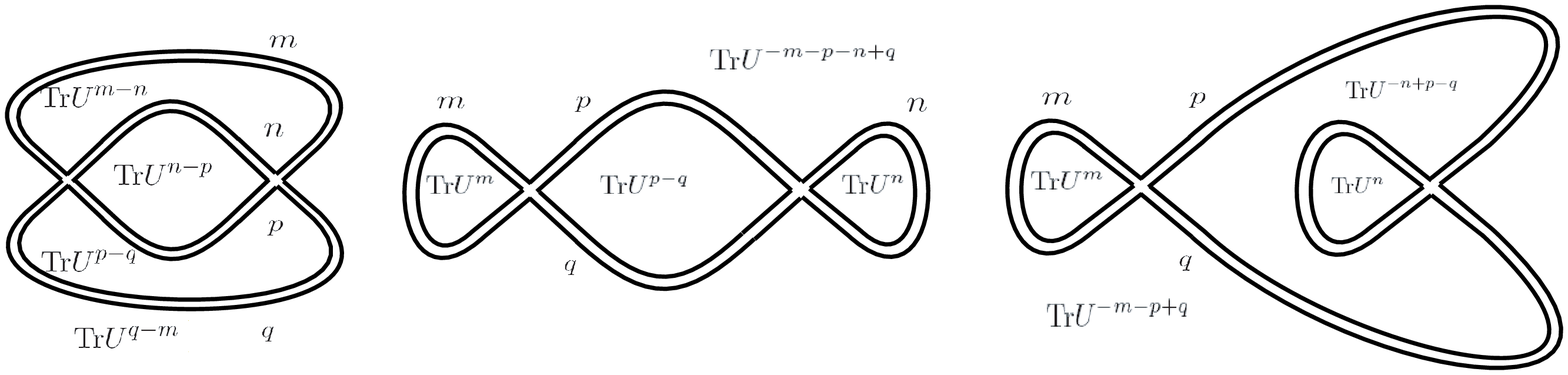} }

\ifig\TorExp{Some non-planar (torus) connected contributions to
$\vev{\Tr M^4(\tau) \Tr M^4(0)}$. For visualization purpose, the
edge of one of the faces is drawn in red.}{\epsfxsize=11cm
\epsfbox{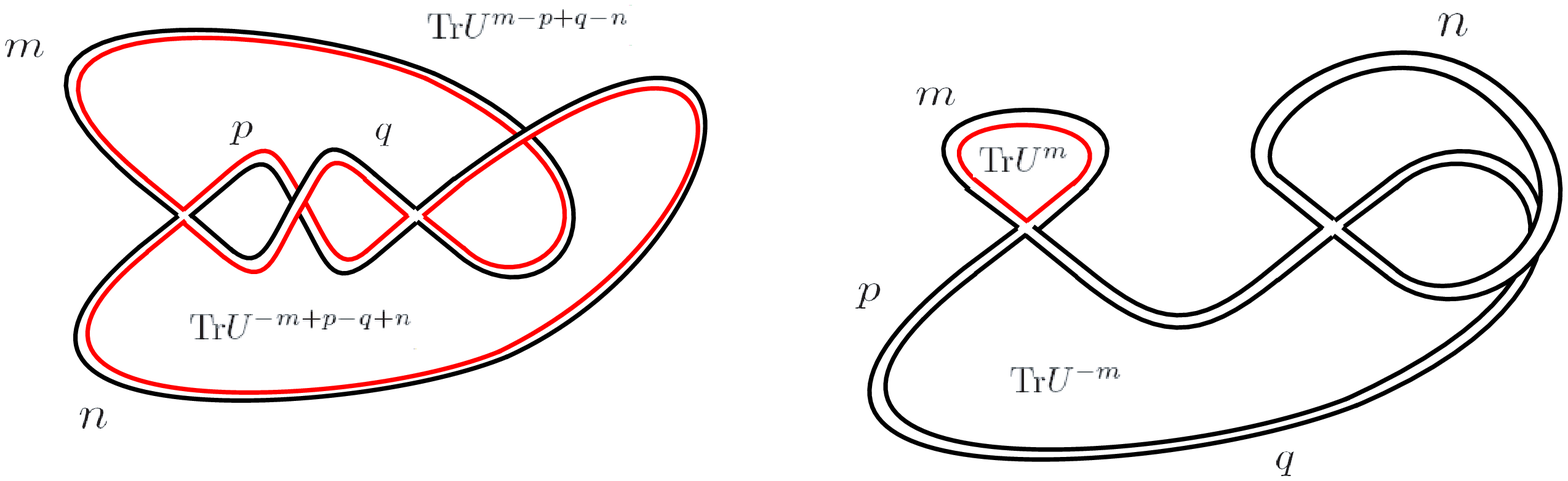} }

 The
connected planar contributions are given in \PlanExp\ with, for
example, the first diagram given by
 \eqn\conttwo{\eqalign{
  {4\ov N^2}\!\! \!\!\!\! \sum_{m,n,p,q=-\infty}^\infty \!\! \!
 \!\!\!\!
 G_s(\tau- m \beta)G_s(\tau- n
\beta)G_s(\tau- p \beta)G_s(\tau- q \beta) \vev{\Tr U^{m-n}\Tr
U^{n-p}\Tr U^{p-q}\Tr U^{q-m}}_U
 }}
In \TorExp\ we have also plotted some connected non-planar
diagrams, with the first diagram given by
 \eqn\contthree{\eqalign{
{4\ov N^2}\!\! \!\!\!\! \sum_{m,n,p,q=-\infty}^\infty \!\! \!
 \!\!\!\!
 G_s(\tau- m \beta)G_s(\tau- n
\beta)G_s(\tau- p \beta)G_s(\tau- q \beta) \vev{ \Tr U^{m-p+q-n }
\Tr U^{-m+p-q+n}}_U
 }}


\ifig\VorDiag{Connected vortex diagram from disconnected double
line diagram }{\epsfxsize=4cm \epsfbox{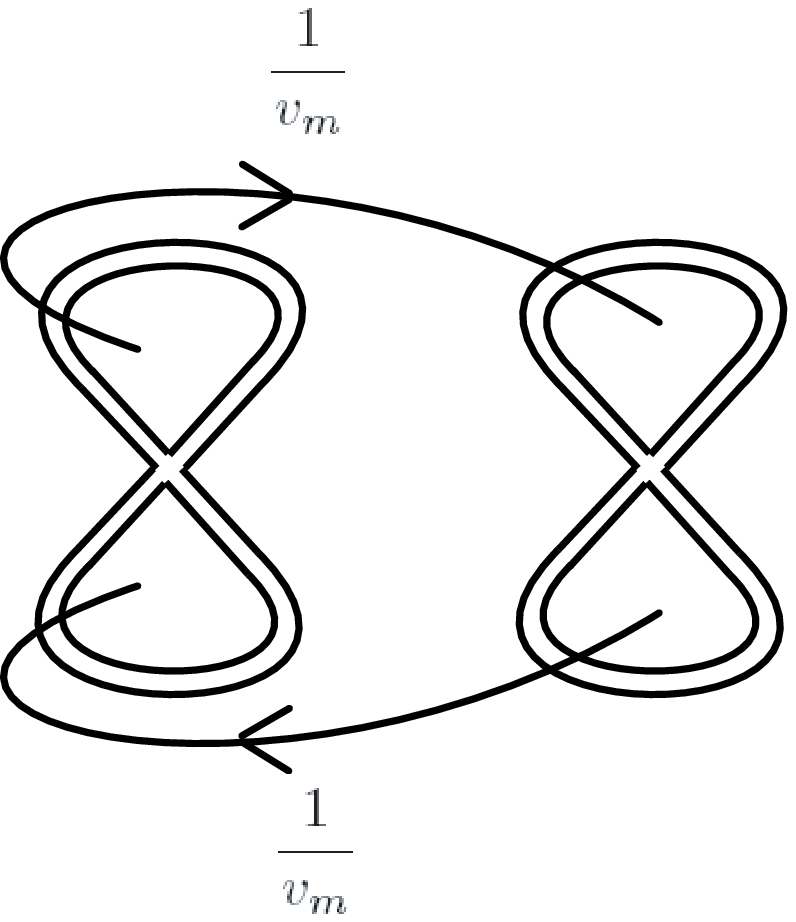} }

Now let us consider the evaluation of the expectation values of
traces of $U$ in \contone-\contthree\ using \rTrU. At leading
order in  the large $N$ expansion the expectation values give
$N^F$, where $F$ is the number of traces, times some product of
Kronecker delta enforcing all exponents to be zero. In this case
we recover the results of~\refs{\brigante}. Higher order
corrections in $1/N$ can be described graphically by inserting
pairs of vortices on different faces of the diagrams and
connecting them with the propagator $1\ov v_b(\beta)$. One should
sum over all the possible ways of inserting pairs of vortices.
Note that each vortex insertion gives a factor of $1/N$. Diagrams
with disconnected parts connected by vortex propagators should be
considered as connected as in~\VorDiag. Note that in terms of
large $N$ counting~\VorDiag\ is of the same order as those in
\TorExp\ with no vortices.

\appendix{B}{Proof of~\corrTrU}

In this appendix we prove equation~\corrTrU. In the next subsection
we discuss some elementary aspects of $U(N)$ group integrals. We
then proceed to evaluate \parT. Equation \corrTrU\ is proved in the
end.

\subsec{Group integrals over $U(N)$}

Consider the following integral over the unitary group $U(N)$
 \eqn\defin{
 I= {1 \ov V_N} \int d U \prod _{i=1}^ k (\Tr U^{a_i})^{b_i} \prod
_{j=1}^ s (TrU^{-c_j})^{d_j}}
 where $a_i, b_i, c_i, d_i$ are positive integers
 and
  \eqn\defD{D=\sum_{i=1} ^k
 a_i b_i=\sum_{j=1}^s c_j d_j \ .
 }
$V_N$ is the volume of $U(N)$.

Products of traces of $U$ can be expanded  in terms of characters
of irreducible representations of $U(N)$, which are in one to one
correspondence with irreducible representations of the symmetric
group (see for example~\refs{\ConstableRC}),
 \eqn\prodexp{\prod
_{i=1}^ k (\Tr U^{a_i})^{b_i} = \sum_\lambda \chi_\lambda (a_i,
b_i )\chi_\lambda(U)
 }
 where $\lambda$ labels the irreducible
representations of the symmetric group $S_D$. $\chi_\lambda
(a_i,b_i)$ is the character of the conjugacy class\foot{Recall
that two elements of $S_D$ are conjugate if and only if they
consist of the same number of disjoint cycles of the same lengths.
Denote the number of cycles of length $a_i$ by $b_i$ then a
conjugacy class in $S_D$ is given by a set of $k$ couples
$\{(a_i,b_i)\}\;\;\;i=1,..k$ such that $\sum_{i=1}^k a_i b_i=D$.}
of $S_D$ given by the set $\{(a_i,b_i)\}$ in the representation
$\lambda$. $\chi_\lambda (U)$ is the character of $U$ in the
irreducible representation of $U(N)$ labelled by $\lambda$. Now by
using the orthogonality property for characters we can write:
 \eqn\rnsm{\eqalign{
 I & = \sum_{\lambda \lambda'} \chi_\lambda (a_i, b_i)\chi_{\lambda'} (c_i, d_i)
 {1 \ov V_N} \int d U \chi_\lambda (U)\chi_{\lambda'}(U^{\dagger}) \cr
  & =  \sum_{\lambda} \chi_\lambda (a_i,
b_i)\chi_{\lambda} (c_i, d_i) \ .
 }}
The evaluation of \rnsm\ can be divided into the following two
cases:

\item{1.} If $D\leq N$, then the sum over $\lambda$ can be
evaluated giving~\refs{\ConstableRC}
 \eqn\ICompl{I=
\delta_{\{(a_i,b_i)\},\{(c_i,d_i)\}} \sum_{\lambda} \chi_\lambda
(a_i, b_i)^2 = \delta_{\{(a_i,b_i)\},\{(c_i,d_i)\}}
 \prod_{i=1} ^k
 {a_i} ^{b_i} b_i !}
where the completeness of characters of the symmetric group $S_D$
enforces the sets $\{(a_i,b_i)\}$ and $\{(c_i,d_i)\}$ to define
the same conjugacy class in $S_D$, i.e., to be the same apart from
reordering. This means that the integral is zero for $D<N$ unless
for any factor of $Tr[U^a]^b$ in the integrand there is a
corresponding factor of $Tr[U^{-a}]^b$.

\item{2.} If $D>N$ one needs to restrict the sum over the
irreducible representations $\lambda$ to the representations where
$\chi_\lambda (U) \neq 0$, that we will indicate formally as
$\lambda<N$. In this case the result is more complicated and we do
not have a closed form expression. For the case in which the sets
$\{(a_i,b_i)\}$ and $\{(c_i,d_i)\}$ are equal up to reordering one
 has
  \eqn\Ired{
  I= \sum_{\lambda<N} \chi_\lambda
(a_i, b_i)^2 <  \prod_{i=1} ^k {a_i} ^{b_i} b_i !
 \ .}

\subsec{Partition function integrals}

We now consider the evaluation of the free theory partition function
\parT. To warm up let us consider the following
integral
 \eqn\caszone{\eqalign{
 {1 \ov V_N}  \int
 d U e^{z_1 \Tr U\Tr U^{\dagger}} & = {1 \ov V_N} \int d U
\sum_{p=0} ^\infty{1\ov p!}  ( z_1 \Tr U\Tr U^{\dagger}) ^p \cr
 & =  \sum_{p=0}^N z_1^p + O(z_1 ^N ) \cr
 & =  {1\ov
 1-z_1}+O(z_1^N) \ .
 }}
For $0 < z_1<1$ the corrections to the $N=\infty$ result are of
order $O(z_1^N)$ and are therefore exponentially suppressed in
$N$. In the more general case \parT\ (with $V_n(\beta)=z_n$) one
can proceed exactly as above, writing
 \eqn\gencase{ \eqalign{
  Z_0 & =  {1 \ov V_N} \int d U \, e^{I_0 (U)} =
  {1 \ov V_N} \int d U \, \exp \le(\sum_{n=1}^\infty {z_n \ov n} \Tr U^n \Tr U^{\dagger n} \ri)
 \cr
  & = {1 \ov V_N} \int d U  \, \prod_{n=1}^\infty \le(
\sum_{p_n=0}^\infty { z_n^{p_n}  \ov p_n! \, n^{p_n}} \left(\Tr
U^n \Tr U^{-n}\right)^{p_n} \ri) \cr
 & = \prod_{n=1} ^\infty {1\ov 1-z_n }-C(N) \
 }}
where $C(N)$ is given by
 \eqn\ejnS{\eqalign{
 C(N) & = \le[\prod_{n=1}^\infty \le(
\sum_{p_n=0}^\infty z_n^{p_n} \ri)  - {1 \ov V_N} \int d U \,
  \prod_{n=1}^\infty \le(
\sum_{p_n=0}^\infty { z_n^{p_n}  \ov p_n! \, n^{p_n}} \left(\Tr
U^n \Tr U^{-n}\right)^{p_n} \ri)\ri]_{\sum_n n p_n > N} \cr
 & < \le[\prod_{n=1}^\infty \le(
\sum_{p_n=0}^\infty z_n^{p_n} \ri) \ri]_{\sum_n n p_n > N}
 \equiv D(N)
 }}
Note the the subscript in the above equation indicates that one
should only sum over those $p_n$ which satisfy $\sum_n n p_n > N$.
$D(N)$ can be estimated as follows. Consider the expansion
 \eqn\sert{\prod_{n=1} ^\infty {1\ov 1-z_n t^n
 }=\sum_{n=0}^{\infty} a_n(z_1,z_2,...) t^n
 }
where $a_n$ are polynomials in the $z_i$ with positive
coefficients. Note that
  \eqn\minor{
   D(N)=  \sum_{n=N+1}^{\infty}
a_n(z_1,z_2,....) \ .
 }
 Define
\eqn\zstar{z_*=\max(z_1,z_2^{\ha},z_3^{1\ov 3},...,z_n^{1\ov
n},...)
 }
Below $T_H$,  we have $z_*<1$. Then we have
$0<a_n(z_1,z_2,...)<a_n(z_*,z_*^2,z_*^3,...)=b_n z_*^n$ where the
$b_n$'s are the coefficients of the series of $ \prod_{m=1}
^\infty {1\ov 1-z_*^m } = \sum_{n=0}^\infty b_n z_*^n$. This
series has radius of convergence equal to $1$ because the function
has no singularities for $|z_*|<1$. It then follows that for a
given $\ep>0$ there exists an $M(\ep)$ such that for $n>M(\ep)$ it
is true that $b_n< (1+\ep)^n$. Then for $\ep< {1\ov z_*}-1$ and
$N>M(\ep)$ the following holds:
 \eqn\corrmaj{
  C(N) < D(N)<\sum_{n=N+1}^{\infty}
((1+\ep)z_*)^n={((1+\ep)z_*)^{N+1}\ov{1-(1+\ep)z_*}}} and
therefore the corrections are exponentially small since
$(1+\ep)z_*<1$.

To summarize we find that
 \eqn\gencaseb{
  Z_0= {1 \ov V_N} \int d U \, \exp \le(\sum_n {z_n \ov n}
\Tr U^n \Tr U^{\dagger n} \ri) = \prod_{n=1} ^\infty {1\ov 1-z_n
}-K e^{-N c}
 }
 where $c=-\log(z^*)>0$ and $K>0$.

 \subsec{Correlation functions}

Correlation functions \usna\
 \eqn\rjjd{
 \vev{ \prod _{i=1}^ k (\Tr U^{a_i})^{b_i} \prod
_{j=1}^ s (TrU^{-c_j})^{d_j}}_U = {1 \ov Z_0} \int dU \, e^{I_0
(U)} \, \prod _{i=1}^ k (\Tr U^{a_i})^{b_i} \prod _{j=1}^ s
(TrU^{-c_j})^{d_j}
 }
where $a_i, b_i, c_i, d_i$ are positive integers of order $O(N^0)$
can now be calculated easily using the technique above.
Correlation functions of the form $\vev{\prod_n (\Tr U^ {a_n} \Tr
U^{- a_n})^{b_n}}_U$ are obtained by taking derivatives on $Z_0$
\gencaseb\ with respect to ${z_n\ov n}$
 \eqn\derZ{
 \vev{\prod_n
(\Tr U^ {a_n} \Tr U^{- a_n})^{b_n}}={1\ov Z_0}\prod_n
n^{b_n}{d^{b_n} Z_0 \ov d z_n^{b_n}} \ .
 }
If in \rjjd\ the $\{(a_i,b_i)\}$ are not matched with
$\{(c_i,d_i)\}$ up to reordering, due to \ICompl, the correlation
function is zero up to nonperturbative corrections which are of
order $(z^*)^N$. For example, $\vev{\Tr U^a \Tr U^a \Tr
U^{-2a}}_U$ is zero at any finite order in ${1\ov N^2}$ expansion
unless $a$ is zero.

The above results can be summarized by the following: the
integrals can be evaluated by treating each $\Tr U^n$ as an
independent integration variable. More explicitly, replacing
 \eqn\fjje{
  {1 \ov N} \Tr U^n \to \phi_n , \qquad {1 \ov N} \Tr U^{-n} \to
  \phi_{-n} =
  \phi_n^*, \qquad \phi_0 =1
 }
then
 \eqn\corrTrU{\eqalign{
  & \vev{{1 \ov N} \Tr U^{s_1} {1 \ov N} \Tr U^{s_2} \cdots {1 \ov N} \Tr U^{s_F}}_U \cr
  & = {1 \ov Z_0} \int_{-\infty}^\infty \prod_{i=1}^\infty d \phi_i d \phi_i^*
  \; \phi_{s_1} \cdots \phi_{s_F} \; \exp \le(- N^2 \sum_{n=1}^\infty
  {1-z_n\ov n} \phi_n \phi_n^* \ri) + {\rm nonperturbative \;\; in} \;
  N \ .
 }}

\listrefs
\end
\end